\documentclass[prd,aps,preprint,tightenlines,superscriptaddress]{revtex4}
\usepackage{epsfig}
\newcommand{\insertfig}[2]{\mbox{\epsfxsize=#1cm \epsfbox{#2.eps}}}
\usepackage{axodraw}
\preprint{DOE/ER/40762-322} \preprint{UM-PP\#05-006}

\begin{document}
\title{Deuteron Compton Scattering in Effective Field Theory:
\\ Spin-Dependent Cross Sections and Asymmetries}
\author{Jiunn-Wei Chen} \email{jwc@phys.ntu.edu.tw} \affiliation{Department of Physics,
 National Taiwan University, Taipei, Taiwan 10617}
\author{Xiangdong Ji} \email{xji@physics.umd.edu} \affiliation{Department of
Physics, University of Maryland, College Park, Maryland 20742}
\author{Yingchuan Li}
\email{yli@physics.umd.edu} \affiliation{Department of Physics,
University of Maryland, College Park, Maryland 20742}
\date{\today}
\vspace{0.5in}
\begin{abstract}

Polarized Compton scattering on the deuteron is studied in nuclear
effective field theory. A set of tensor structures is introduced
to define 12 independent Compton amplitudes. The scalar and vector
amplitudes are calculated up to ${\cal O}((Q/\Lambda)^2)$ in
low-energy power counting. Significant contribution to the vector
amplitudes is found to come from the spin-orbit type of
relativistic corrections. A double-helicity dependent cross
section $\Delta_1 \sigma =
\left(\sigma_{+1-1}-\sigma_{+1+1}\right)/2$ is calculated to the
same order, and the effect of the nucleon isoscalar spin-dependent
polarizabilities is found to be smaller than the effect of
isoscalar spin-independent ones. Contributions of spin-independent
polarizabilities are investigated in various asymmetries, one of
which has as large as 12 (26) percent effect at the center-of-mass
photon energy 30 (50) MeV.
\end{abstract}

\maketitle

\section{Introduction}
Compton scattering is an important tool to probe the internal
structure of a composite system, such as an atomic nuclei. As
quantum electrodynamics involved in the process is well
understood, the remaining uncertainty is associated with the
strong interactions among nucleons in nuclei. Thus Compton
scattering data enables physicists to extract information about
the nuclear structure from the underlying strong interaction
dynamics. Recent progress in high-energy, high-intensity photon
beams has made Compton scattering a practical tool for nuclear
physicists \cite{higs}. In particular, a polarized photon beam is
capable of studying spin aspects of strong interaction physics.
This paper focuses on polarized Compton scattering on the
deuteron, the double-helicity dependent cross section in
particular, in the framework of nuclear effective field theory
(EFT).

The deuteron, as the simplest nuclear system, is of great
importance for understanding the nucleon-nucleon interactions and
the properties of individual nucleons. Polarized Compton
scattering on the deuteron presents a new opportunity to probe
spin physics. Indeed, because the deuteron is a loosely bound
system, one might expect to learn a host of spin-dependent
properties of the neutron and proton as free particles. This
possibility is especially important for the structure of the
neutron because there is no free neutron target in nature.

It has been realized for some time that nuclear physics at low
energy might be understood by effective field theories (EFT) which
works according to the same principles as the standard model
\cite{weinberg}. However, constructing a workable scheme for
specific systems is not straightforward. In the past few years,
considerable progress has been made in the two-nucleon sector (see
\cite{review} for a recent review). It began with the pioneering
work of Weinberg, who proposed to encode the short-distance
physics in a derivative expansion of local operators
\cite{weinberg}. The problem associated with the unusually small
binding energy of the deuteron was solved by Kaplan, Savage and
Wise by exploiting the freedom of choosing a renormalization
substraction scheme \cite{kaplan1}, quickly followed by the
pionless version \cite{chen1} (see also
\cite{vK97,Cohen97,BHvK1}). Required reproducing the residue of
the deuteron pole at next-to-leading order (NLO), a version with
accelerated convergence was suggested in \cite{phillips}. The use
of dibaryon fields as auxiliary fields, first introduced in
\cite{kaplan2}, was taken seriously in \cite{savage} which
simplified the calculation significantly.

From the viewpoint of nuclear EFT,  Compton scattering on the
deuteron at low energy can be divided into two regions according
to the photon energy $\omega$. Region I is where the photon energy
is far below the binding energy of deuteron $B = 2.2$ MeV and,
hence, $\omega/B$ is a small parameter. Region II is where the
photon energy is above the binding energy, but significantly below
the mass of the pion, for example, $\omega \sim$ 50 MeV. In Region
I, one makes the low energy expansion of Compton amplitudes and
studies various polarizabilities of the deuteron defined through
the expansion \cite{Ji}. Studies in this ultra-low energy region,
where the binding effect plays a dominant role, provide insight
about the internal structure of the deuteron as a bound state. In
Region II, the probing photon is more sensitive to the responses
from individual nucleons. Therefore, Compton scattering there may
serve as an alternative tool to study free-nucleon properties,
such as spin-independent and dependent polarizabilities. In this
paper, we are mostly interested in the second region.

Extracting the isoscalar spin-independent polarizabilities
$\alpha_0$ and $\beta_0$ from unpolarized Compton scattering has
attracted considerable attention in the past two decades. Although
there are three types of amplitudes (scalar, vector and tensor)
contributing to the cross section, only the scalar amplitudes have
been included in some of the calculations of the unpolarized cross
section so far. Nuclear EFT seems to provide a justification for
this. However, a recent work \cite{unpolarized} showed that vector
amplitudes contribute significantly (order of 15\% or more) to the
unpolarized cross section, because of the enhancement from a
factor of the square of the isovector magnetic moment $\mu^2_1$.
It turns out that this enhancement has its effect not only on
unpolarized scattering but also on polarized one, leading to, for
instance, a bigger helicity-dependent cross section. Although this
makes it easier to measure it experimentally, the effect also
diminishes the contribution from the isoscalar nucleon
spin-dependent polarizabilities, and hence makes it harder to
access them from the future Compton data.

To demonstrate the above point, we calculate a double-helicity
dependent (vector-polarized) cross section up to the order at
which the spin polarizabilities contribute, and compare the
results with and without their contribution. The photon-nucleon
interactions considered in this calculation include the electric
current and magnetic couplings, and the spin-orbit terms from the
non-relativistic reduction of the relativistic interactions. It
has been realized previously that the relativistic corrections are
surprisingly large in potential model calculations
\cite{weyrauch,wilbois,levchuk,lvov,karakowski}. In EFT, the
spin-orbit interactions were taken into account in the studies of
the deuteron forward spin-dependent polarizabilities \cite{Ji} and
Drell-Hearn-Gerasimov sum rule \cite{DHG}. They were neglected in
other EFT calculations because they are nominally suppressed in
power counting by $1/M_N$ relative to the other two couplings.
However, for certain spin-dependent observables, their
contributions can be of leading order, as we shall see.

The paper is organized as follows. Section II is devoted to
kinematics, where we write down 12 basis structures for scattering
amplitudes using parity and time-reversal symmetries. The scalar
and vector structures are the same as those in Compton scattering
on a spin-1/2 particle such as the proton. The tensor structures
are new, and useful for general discussions of polarized deuteron
Compton scattering. Section III explains a calculation of the
vector Compton amplitudes using the dibaryon formulation of EFT.
Power counting in both Regions I and II is explained to show the
significant contribution of the spin-orbit interactions. The
result of individual diagrams is listed in Appendix B. In Sec. IV,
a double-helicity-dependent (vector-polarized) cross section is
defined, and the numerical result is presented with and without
the contribution from the nucleon spin-dependent polarizabilities.
The feasibility of using polarized Compton data to extract these
polarizabilities is discussed. In Sec. V, we investigate the
effect of the spin-independent polarizabilities on a number of
spin asymmetries. Section VI contains the conclusion of the paper.

\section{Real Photon-Deuteron Compton Scattering
 Amplitudes}

In this section, the general tensor structure of the amplitudes
for real photon Compton scattering on a deuteron is considered.
Through helicity counting, it is easy to see that there are a
total of 12 independent amplitudes. We choose these amplitudes in
a basis convenient for subsequent calculations. We comment on the
frame dependence of the tensor structures associated with the
amplitudes.

The real photon has two independent helicities $\pm 1$; the
deuteron has three, $\pm 1$ and 0. Therefore, the total number of
helicity amplitudes is $2\times 3\times 2\times 3=36$. Parity
invariance of strong and electromagnetic interactions restricts
the number of independent ones to 36/2=18. Among those,
time-reversal symmetry relates 6 to the others with initial and
final state exchanged. This reduces the number of independent
amplitudes to $18-6=12$. Moreover, the general result of helicity
counting can be derived, and is $2(J+1)(2J+1)$ for a spin-$J$
target.

In the low-energy region, it is convenient to use the
nonrelativistic notation for tensor structures associated with the
amplitudes. If the spins of the initial and final deuterons are
coupled, the sum is 0, 1 or 2. The amplitudes classified in this
way are called scalar, vector, and tensor, respectively. Clearly
the number of scalar amplitudes must be the same as that of
Compton scattering amplitudes on a spin-0 target, namely, 2; and
the number of vector amplitudes is the same as that on a spin-1/2
target, 4. Thus the number of independent tensor amplitudes is
$12-2-4=6$.

In the remainder of this section, we construct a set of 12
linearly-independent structures, using the 3-momenta of the photon
and deuteron, and their polarization vectors. Among four
3-momenta, only three are independent because of the momentum
conservation. By choosing a specific frame, one more constraint
follows, and hence only initial and final 3-momenta of photon,
$\vec{k}$ and $\vec{k}'$, are needed for the construction. The
initial and final three-momenta of the deuteron, $\vec{p}$ and
$\vec{p}'$ can be expressed in terms of these of the photon. For
example, the lab frame is defined by $\vec{p}=0$ and
$\vec{p}'=\vec{k}-\vec{k}'$, the center-of-mass frame (CM) by
$\vec{p}=-\vec{k}$ and $\vec{p}'=-\vec{k}'$, and the so-called
Breit frame by $\vec{p}=\frac{1}{2}(\vec{k}'-\vec{k})$ and
$\vec{p}'=-\frac{1}{2}(\vec{k}'-\vec{k})$ and so $\vec{p} +
\vec{p}'=0$. The constraints among momenta associated with a frame
are generally not invariant under symmetries such as time
reversal, which exchanges the initial and final momenta and
reverses their directions, and photon crossing symmetry, which
exchanges the initial and final photon with the sign of energy and
3-momentum flipped. For instance, the momentum constraint in the
lab frame is not invariant under either time reversal or crossing
symmetry, while the momentum constraint in the CM frame violates
crossing symmetry.

According to the above, in the CM and Breit frames where parity
and time reversal invariance are manifest, there are 12
independent tensor structures for the Compton amplitudes. These
structures are constructed out of initial and final photon
polarization vectors ($\hat{\epsilon}'^{*}$ and $\hat{\epsilon}$),
deuteron polarization vectors ($\hat{\xi}'^{*}$ and $\hat{\xi}$),
and the initial and final photon momentum vectors
$\hat{k}={\vec{k}}/{|\vec{k}|}$ and
$\hat{k}'={\vec{k}'}/{|\vec{k}'|}$. One can couple
($\hat{\xi}'^{*}$ and $\hat{\xi}$) into scalar, vector and tensor
to obtain scalar, vector, and tensor amplitudes. Alternatively,
these structures can be obtained by the matrix element of a unit
matrix $I$, spin matrices $J_i$, or tensor $(J_iJ_j+J_jJ_i-{\rm
trace})$ between the initial and final deuteron polarization
vectors. Under parity transformation, all momentum and
polarization vectors change sign, whereas the spin matrices do
not. Under time-reversal transformation, these quantities
transform according to: $ \hat{\epsilon} \Leftrightarrow
\hat{\epsilon}'^{*}; \*\* \hat{k}\Leftrightarrow-\hat{k}'; \vec{J}
\Rightarrow -\vec{J}$.

Requiring symmetry under both parity and time-reversal, we choose
the 12 basis structures for Compton scattering on the deuteron as
follows,
\begin{eqnarray}
\rho_1 &=&  \hat{\epsilon}'^\ast \cdot \hat{\epsilon} \; I =
\epsilon'^\ast \cdot \hat{\epsilon} \; \xi '^\ast\cdot \xi ~, \nonumber \\
\rho_2 &=& \hat{s}'^{*}\cdot \hat{s} \; I = \hat{s}'^{*}\cdot
\hat{s} \; \xi
'^\ast\cdot\xi ~,  \nonumber \\
\rho_3 &=& i  J \cdot \hat{\epsilon}'^\ast \times \hat{\epsilon} =
(\xi'^{\ast} \times
\xi) \cdot (\hat{\epsilon}'^{\ast} \times \hat{\epsilon}) ~, \nonumber \\
\rho_4 &=& i J \cdot \hat{s}'^{*} \times \hat{s} = (\xi'^{\ast} \times \xi) \cdot (\hat{s}'^{*} \times \hat{s}) ~, \nonumber \\
\rho_5 &=& i (J \cdot \hat{k} \hat{s}'^{*} \cdot \hat{\epsilon} -
J \cdot \hat{k}' \hat{\epsilon}'^\ast\cdot \hat{s}) =
[(\xi'^{\ast} \times \xi) \cdot \hat{k} \hat{s}'^{*} \cdot
\hat{\epsilon} - (\xi'^{\ast} \times \xi) \cdot
\hat{k}'\hat{\epsilon}'^{\ast} \cdot \hat{s}] ~, \nonumber \\
\rho_6 &=& i (J \cdot \hat{k}' \hat{s}'^{*} \cdot \hat{\epsilon} -
J \cdot \hat{k} \hat{\epsilon}'^\ast\cdot \hat{s}) =
[(\hat{\xi}'^{*} \times \hat{\xi}) \cdot \hat{k}' \hat{s}'^{*}
\cdot \hat{\epsilon} - (\xi'^{\ast} \times \xi) \cdot
\hat{k}\hat{\epsilon}'^{\ast} \cdot \hat{s}] ~, \nonumber \\
\rho_7&=&-(J \cdot \hat{\epsilon} J \cdot \hat{\epsilon}'^{*} + J
\cdot \hat{\epsilon}'^{*} J \cdot
\hat{\epsilon}-\frac{4}{3}\hat{\epsilon}'^{*} \cdot
\hat{\epsilon})=\hat{\xi}'^{*} \cdot \hat{\epsilon}'^{*} \hat{\xi}
\cdot \hat{\epsilon} + \hat{\xi}'^{*} \cdot \hat{\epsilon}
\hat{\xi} \cdot \hat{\epsilon}'^{*} -\frac{2}{3}\hat{\xi}'^{*}
\cdot
\hat{\xi}\hat{\epsilon}'^{*} \cdot \hat{\epsilon} ~, \nonumber \\
\rho_8&=&-(J \cdot \hat{s} J \cdot \hat{s}'^{*} + J \cdot
\hat{s}'^{*} J \cdot \hat{s}-\frac{4}{3}\hat{s}'^{*} \cdot
\hat{s})=\hat{\xi}'^{*} \cdot \hat{s}'^{*} \hat{\xi} \cdot \hat{s}
+ \hat{\xi}'^{*} \cdot \hat{s} \hat{\xi} \cdot \hat{s}'^{*}
-\frac{2}{3}\hat{\xi}'^{*} \cdot
\hat{\xi}\hat{s}'^{*} \cdot \hat{s} ~, \nonumber \\
\rho_9&=&-\hat{\epsilon}'^{*} \cdot \hat{k} (J \cdot
\hat{\epsilon} J \cdot \hat{k}' + J \cdot \hat{k}' J \cdot
\hat{\epsilon})-\hat{\epsilon} \cdot \hat{k}' (J \cdot
\hat{\epsilon}'^{*} J \cdot \hat{k} + J \cdot
\hat{k} J \cdot \hat{\epsilon}'^{*})+\frac{8}{3}\hat{\epsilon}'^{*} \cdot \hat{k} \hat{\epsilon} \cdot \hat{k}' \nonumber \\
&=&\hat{\epsilon}'^{*} \cdot \hat{k} (\hat{\xi}'^{*} \cdot
\hat{k}' \hat{\xi} \cdot \hat{\epsilon}
 + \hat{\xi}'^{*} \cdot \hat{\epsilon} \hat{\xi} \cdot \hat{k}') + \hat{\epsilon} \cdot \hat{k}'
 (\hat{\xi}'^{*} \cdot \hat{k} \hat{\xi} \cdot \hat{\epsilon}'^{*}
 + \hat{\xi}'^{*} \cdot \hat{\epsilon}'^{*} \hat{\xi} \cdot \hat{k}) -
 \frac{4}{3}\hat{\xi}'^{*} \cdot \hat{\xi} \hat{\epsilon}'^{*} \cdot \hat{k} \hat{\epsilon} \cdot \hat{k}' ~, \nonumber \\
\rho_{10}&=&-\hat{s}'^{*} \cdot \hat{k} (J \cdot \hat{s} J \cdot
\hat{k}' + J \cdot \hat{k}' J \cdot \hat{s})-\hat{s} \cdot
\hat{k}' (J \cdot \hat{s}'^{*} J \cdot \hat{k} + J \cdot
\hat{k} J \cdot \hat{s}'^{*})+\frac{8}{3}\hat{s}'^{*} \cdot \hat{k} \hat{s} \cdot \hat{k}' \nonumber \\
&=&\hat{s}'^{*} \cdot \hat{k} (\hat{\xi}'^{*} \cdot \hat{k}'
\hat{\xi} \cdot \hat{s}
 + \hat{\xi}'^{*} \cdot \hat{s} \hat{\xi} \cdot \hat{k}') + \hat{s} \cdot \hat{k}'
 (\hat{\xi}'^{*} \cdot \hat{k} \hat{\xi} \cdot \hat{s}'^{*}
 + \hat{\xi}'^{*} \cdot \hat{s}'^{*} \hat{\xi} \cdot \hat{k}) -
 \frac{4}{3}\hat{\xi}'^{*} \cdot \hat{\xi} \hat{s}'^{*} \cdot \hat{k} \hat{s} \cdot \hat{k}' ~, \nonumber \\
\rho_{11}&=&-\hat{\epsilon}'^{*} \cdot \hat{\epsilon} (J \cdot
\hat{k} J \cdot \hat{k} + J \cdot \hat{k}' J \cdot \hat{k}'
-\frac{4}{3})=\hat{\epsilon}'^{*} \cdot
\hat{\epsilon}(\hat{\xi}'^{*} \cdot \hat{k}\hat{\xi} \cdot \hat{k}
+ \hat{\xi}'^{*} \cdot \hat{k}'\hat{\xi} \cdot \hat{k}'
-\frac{2}{3} \hat{\xi}'^{*} \cdot \hat{\xi}) ~,
\nonumber \\
\rho_{12}&=&-\hat{s}'^{*} \cdot \hat{s} (J \cdot \hat{k} J \cdot
\hat{k} + J \cdot \hat{k}' J \cdot \hat{k}'
-\frac{4}{3})=\hat{s}'^{*} \cdot \hat{s}(\hat{\xi}'^{*} \cdot
\hat{k}\hat{\xi} \cdot \hat{k} + \hat{\xi}'^{*} \cdot
\hat{k}'\hat{\xi} \cdot \hat{k}' - \frac{2}{3}\hat{\xi}'^{*} \cdot
\hat{\xi}) ~ ,
\end{eqnarray}
where the $\hat{s}$ and $\hat{s}'^{*}$ are defined as $\hat{s} =
\hat{k} \times \hat{\epsilon}$ and $\hat{s}'^{*} = \hat{k}' \times
\hat{\epsilon}'^{*}$. These structures are constructed in such a
way that duality between the electric and magnetic fields is
manifest. Under the dual transformation, $\hat{\epsilon}
\Rightarrow \hat{s}$, $\hat{s} \Rightarrow -\hat{\epsilon}$, which
is a $\pi/2$-rotation in the photon polarization, the above
structures transform as $\rho_{2i-1} \Leftrightarrow \rho_{2i}$
with $i=1,...,6$. The structures with the unit matrix and spin
operators ($\rho_1$ to $\rho_6$) are the same as those for a
spin-1/2 target \cite{babusci}. Appendix A explains why these 12
structures are complete and independent.

The most general Compton scattering amplitude on the deuteron can
be expressed as
\begin{equation}
f=\sum_{i=1}^{12}f_i\rho_i \ ,
\end{equation}
where $f_i$ defines the spin-dependent amplitudes. The first two
($i=1, 2$) are scalar amplitudes; the following four ($i=3,...,
6$) are vector amplitudes; and the last six ($i=7,...,12$) are
tensor amplitudes.

\section{Vector Compton Amplitudes to $(Q/\Lambda)^4$ from EFT}

In this section, we calculate the vector Compton amplitudes to
$(Q/\Lambda)^4$ in a low-energy expansion in nuclear EFT. The
calculation is based on the dibaryon approach in the pionless
theory, which has been referred as dEFT($\not\! \pi$)
\cite{savage}.

A central concept in EFT is power counting. EFT is designed to
describe physics at one scale---low-energy scale in this
case---using an effective lagrangian, and the physics at other
scales is accounted for through the couplings. Power counting
allows a systematic way to take into account corrections from
other energy scales. For Compton scattering on the deuteron, the
natural momentum scale is $\sqrt{M_NB}$ ($M_N$ is the nucleon
mass) which will be generically referred to as $Q$. The deuteron
binding energy $B$ is then counted as order of $Q^2$. The energy
and momentum of the external photon probe, $\omega=|\vec{k}|$, is
counted as
\begin{itemize}
\item{ $Q^2$ in Region I where $\omega\ll B$}, and as \item{$Q$ in
Region II, where $\omega \sim \sqrt{M_N B}$.}
\end{itemize}
The high-energy scales include the nucleon mass $M_N$, the pion
mass $m_\pi$, and similar scales describing the structure of the
nucleon, like the charge radius, and parameters in nucleon-nucleon
interactions. Because $m_\pi$ and $M_N$ are very different, we use
$\Lambda$ to denote scales at around $m_\pi$, and identify
$m_\pi/M_N$ as $Q/\Lambda$. Therefore, ratio $Q/M_N$ can actually
be treated as $(Q/\Lambda)^2$. Although this is not fully
consistent in the EFT sense, it is a way to organize
numerically-close ratios phenomenologically \cite{rupak}.

In dEFT($\not\! \pi$), the nucleon rescattering in both singlet
$^1S_0$ and triplet $^3S_1$ channels is represented by the
propagation of dibaryon fields, $t_j$ and $s_a$, respectively. The
lagrangian density for the triplet channel is \cite{savage}:
\begin{equation}
{\cal L}=N^\dag \left[ i\partial_0 + \frac{\mathbf{D}^2}{2M_N}
\right] N - t^\dag_j \left[ i
\partial_0 + \frac{\mathbf{D}^2}{4M_N}-\Delta \right] t^j - y
\left[t^\dag_j N^T P^j N + h. c.\right] ~ , \label{basic}
\end{equation}
where $N$ is the 2-component nucleon field with an implicit
isospin index. The time and spatial derivatives with
electromagnetic gauge symmetry are $D_0$ and $\mathbf{D}$,
respectively. $P^j=\frac{1}{\sqrt{8}}\tau_2\sigma_2\sigma_j$ is
the projection operator of the triplet channel, and $y$ is the
coupling between nucleons in the triplet channel and the triplet
dibaryon. Requiring producing the nucleon-nucleon scattering
amplitude, one has
\begin{equation}
y^2=\frac{8\pi}{M^2_N r^{(3S_1)}} ,  ~~~   \Delta = \frac{2}{M_N
r^{(3S_1)}} \left(\frac{1}{a^{(3S_1)}}-\mu\right) ~,
\end{equation}
with $\mu$ being the renormalization-scale introduced in the power
divergent subtraction scheme \cite{kaplan1}. The parameters $a$
and $r$ are the scattering length and effective range,
respectively. In the present formulation, these two are counted as
order $Q^{-1}$ in both singlet and triplet channels. Thus the
scaling property of $y$ and $\Delta$ is $y\sim\sqrt{Q}$ and
$\Delta \sim Q^2$, respectively. Dressing the dibaryon propagator
with nucleon bubbles does not change the counting of the
propagator. Therefore the bubbles must be summed to all orders;
the dibaryon propagator dressed with nucleon bubbles is
\begin{equation}
D^{(3S_1)}\left( \overline{E} \right)=\frac{4\pi}{M_N
y^2}\frac{i}{\mu+\frac{4\pi}{M_N
y^2}\left(\Delta-\overline{E}\right)+i\sqrt{M_N \overline{E}}} \ ,
\end{equation}
with $\overline{E}$ the center-of-mass energy. The wave function
renormalization constant is the residue at pole $\overline{E}=-B$,
and a simple calculation yields \cite{savage}: $z_d={\gamma
r^{(3S_1)}}/(1-\gamma r^{(3S_1)})$ \ .

We remark that it is straightforward to convert the nuclear EFT
lagrangian with the nucleon field into that in dEFT($\not\! \pi$).
 Following the
prescription in \cite{savage}, one converts a pair of nucleon
fields in the singlet and triplet channels to dibaryon fields,
\begin{equation}
N^T P^j N \rightarrow \frac{1}{\sqrt{M_N r^{(3S_1)}}} t^j, ~~~ N^T
\overline{P}^a N \rightarrow \frac{1}{\sqrt{M_N r^{(1S_0)}}} s^a,
\end{equation}
where $\overline{P}^a=\frac{1}{\sqrt{8}}\sigma_2\tau_2\tau_a$ is
the projection operator for the singlet channel.

Nuclear EFT describes the interactions between the nucleons and
external electromagnetic probes systematically. Besides the
coupling generated in the covariant derivatives in the above
lagrangian density, $\mathbf{D} = \vec{\nabla}-ie\mathbf{A}$,
there is also the magnetic coupling to the nucleon,
\begin{equation}
{\cal L}_{\rm B} = \frac{e}{2M_N}N^\dagger(\mu_0+
  \mu_1\tau_3 )\mathbf{\sigma}\cdot \mathbf{B} N \ ,  \label{magnetic}
\end{equation}
where $\mu_0$ and $\mu_1$ are the nucleon's isoscalar and
isovector magnetic moments, respectively.  An associated term is
the spin-orbit-type relativistic correction
\begin{equation}
{\cal L}_{\rm SO} = i \frac{e}{8M^2_N} \left( \left(
2\mu_0-\frac{1}{2} \right) + \left( 2\mu_1-\frac{1}{2}
\right)\tau_3 \right) N^\dagger \vec{\sigma}\cdot(\mathbf{D}
\times \mathbf{E} - \mathbf{E} \times \mathbf{D})N \ , \label{SO}
\end{equation}
which is generated from the reduction of a relevant relativistic
interaction.

There are also interaction terms involving the dibaryon fields
themselves. One term accounts for the transition between the
$^3S_1$ and $^1S_0$ channels through a magnetic field,
\begin{equation}
\mathcal{L}_{\mathrm{em, 1}} = e\frac{L_{1}}{M_{N}\sqrt{%
r^{(^{1}S_{0})}r^{(^{3}S_{1})}}}t_{j}^{\dagger
}s_{3}B_{j}+\mathrm{h.c.} ~ . \label{L1}
\end{equation}
The coupling constant $L_1$ has been determined by the rate of $%
n+p\rightarrow d+\gamma$. The measured cross section $\sigma =
334.2\pm 0.5$ mb with an incident neutron speed of $2200$ m/s
fixes $L_1=-4.42$ fm. Another term involves the elastic scattering
of the deuteron in the magnetic field,
\begin{equation}
\mathcal{L}_{\mathrm{em,2}} = -i\frac{e}{M_{N}}\left( \mu_0
-\frac{L_{2}}{r^{(^{3}S_{1})}}\right) \varepsilon
^{ijk}t_{i}^{\dagger }B_{j}t_{k} \ ,  \label{L2}
\end{equation}
with the value of $L_2$ fixed to be $-0.03$ fm from the magnetic
moment of the deuteron. The $\mu_0$ in the above equation is
introduced to reproduce the magnetic moment at leading order
\cite{detmold,ando}. There is also an associated relativistic
correction,
\begin{equation}
\mathcal{L}_{\mathrm{em,2}}^{\mathrm{SO}}=\frac{e}{2M_{N}^{2}}\left(
\mu_0-\frac{L_{2}}{r^{(^{3}S_{1})}}-\frac{1}{4}\right) \varepsilon
^{ijk}t_{i}^{\dagger }\left( \mathbf{D}\times
\mathbf{E}-\mathbf{E}\times \mathbf{D}\right) _{j}t_{k}\ ,
\label{dibaryon SO}
\end{equation}
which generates a seagull interaction of the dibaryon and
electromagnetic fields. At last, there are nucleon
polarizabilities interactions:
\begin{eqnarray}
\mathcal{L}_{\mathrm{pol}}&=& 2\pi N^{\dagger}\left(\alpha_0 +
\alpha_1 \tau_3 \right) \mathbf{E}^2 N + 2\pi
N^{\dagger}\left(\beta_0 + \beta_1 \tau_3 \right) \mathbf{B}^2 N  \nonumber \\
&&  +2\pi N^{\dagger}\left(\gamma^{(s)}_{E1} + \gamma^{(v)}_{E1}
\tau_3 \right) \sigma \cdot \mathbf{E} \times  \dot{\mathbf{E}}  N
+ 2\pi N^{\dagger}\left(\gamma^{(s)}_{M1} + \gamma^{(v)}_{M1}
\tau_3
\right) \sigma \cdot \mathbf{B} \times  \dot{\mathbf{B}}  N  \nonumber \\
&& -2\pi N^{\dagger}\left(\gamma^{(s)}_{E2} + \gamma^{(v)}_{E2}
\tau_3 \right) E_{ij} \sigma_i \mathbf{B}_j  N + 2\pi
N^{\dagger}\left(\gamma^{(s)}_{M2} + \gamma^{(v)}_{M2} \tau_3
\right) B_{ij} \sigma_i \mathbf{E}_j  N ~,  \label{pol}
\end{eqnarray}
where $E_{ij} = 1/2 ( \nabla_i \mathbf{E}_j + \nabla_j
\mathbf{E}_i)$ and $B_{ij} = 1/2 ( \nabla_i \mathbf{B}_j +
\nabla_j \mathbf{B}_i)$ are the electric and magnetic field
gradients. The nucleon isoscalar
($\alpha_0$,$\beta_0$,$\gamma^{(s)}_{E1,M1,E2,M2}$) and isovector
($\alpha_1$,$\beta_1$,$\gamma^{(v)}_{E1,M1,E2,M2}$)
polarizabilities are defined as, for example,
$\alpha_0=1/2(\alpha_p+\alpha_n)$ and
$\alpha_1=1/2(\alpha_p-\alpha_n)$, with similar relations for
others. The isoscalar ones are what can be probed in deuteron
Compton scattering. Chiral perturbation theory calculations yield
\cite{GHM}:
\begin{eqnarray}
&&\alpha_0 = 10\beta_0= 12 \times 10^{-4}\rm{fm}^3, \nonumber \\
&& \gamma^{(s)}_{E1}=-3.1 \times 10^{-4}\rm{fm}^4, ~~
\gamma^{(s)}_{M1}=0.4 \times 10^{-4}\rm{fm}^4,
\nonumber \\
&&\gamma^{(s)}_{E2}=2.1 \times 10^{-4}\rm{fm}^4, ~~
\gamma^{(s)}_{M2}=0.6 \times 10^{-4}\rm{fm}^4 . \label{pol value}
\end{eqnarray}

Feynman diagrams that contribute to the deuteron Compton
scattering to $(Q/\Lambda)^4$ in power counting are shown in Figs.
1-4. Figure 1 contains diagrams with direct photon-dibaryon
interactions. Figure 2 contains the seagull interactions with the
nucleon. The diagram 2c actually corresponds to the contribution
from electromagnetic polarizabilities of the nucleon. Figure 3
include diagrams without intermediate dibaryon fields. Finally
diagrams in Fig. 4 have intermediate singlet and triplet dibaryon
propagations.

%%%%%%%%%%%%%%%%%%%%%%%%%%%%%%%%%
%% FIGURE 1
%%%%%%%%%%%%%%%%%%%%%%%%%%%%%%%%%

\begin{figure}[tbp]
\includegraphics[width=3.6 in]{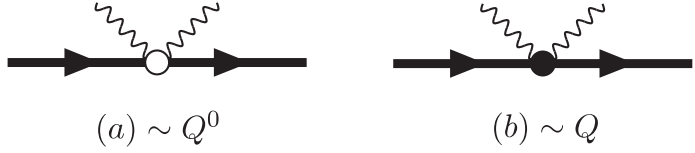}
\caption{\it Compton scattering with photons coupled to the
dibaryon field directly. The open circle denotes the electric
photon-dibaryon coupling from the gauged derivative. The solid dot
denotes the seagull term in Eq. (\protect\ref{dibaryon SO}). The
intermediate thick line represents the triplet dibaryon. }
\label{fig:one}
\end{figure}

\begin{figure}
\SetWidth{0.7}
\begin{center}
\begin{picture}(420,60)(0,0)
\ArrowLine(20,0)(60,0) \SetWidth{3} \ArrowLine(64,15)(80,15)
\ArrowLine(0,15)(20,15) \SetWidth{0.7} \ArrowLine(20,30)(40,30)
\ArrowLine(40,30)(60,30) \Photon(40,30)(20,45){-2}{5}
\Photon(40,30)(58,45){2}{5} \GCirc(40,30){2}{1}
\GOval(20,15)(15,4)(0){0.7} \GOval(60,15)(15,4)(0){0.7}
\Text(40,-10)[]{$(a) \sim Q^0$} \ArrowLine(185,0)(225,0)
\SetWidth{3} \ArrowLine(229,15)(245,15) \ArrowLine(165,15)(185,15)
\SetWidth{0.7} \ArrowLine(185,30)(205,30)
\ArrowLine(205,30)(225,30) \Photon(205,30)(185,45){-2}{5}
\Photon(203,30)(223,45){2}{5} \GCirc(205,30){2}{0}
\GOval(185,15)(15,4)(0){0.7} \GOval(225,15)(15,4)(0){0.7}
\Text(205,-10)[]{$(b) \sim Q$} \ArrowLine(350,0)(390,0)
\SetWidth{3} \ArrowLine(394,15)(410,15) \ArrowLine(330,15)(350,15)
\SetWidth{0.7} \ArrowLine(350,30)(370,30)
\ArrowLine(370,30)(390,30) \Photon(370,30)(350,45){-2}{5}
\Photon(370,30)(388,45){2}{5} \GBox(368,28)(372,32){1}
\GOval(350,15)(15,4)(0){0.7} \GOval(390,15)(15,4)(0){0.7}
\Text(370,-10)[]{$(c) \sim Q, Q^2$}
\end{picture}
\end{center}
\caption{\it Diagrams with seagull interactions on the nucleon
lines. The small open circle denotes the coupling from the gauged
derivative in the first term in Eq. (\ref{basic}), the small solid
circle represents the coupling from spin-orbit interaction defined
in Eq. (\ref{SO}), while the small open box represents the point
interactions associated with polarizabilities of the nucleon in
Eq. (\ref{pol}). Power counting of the leading contribution of
each diagram is listed below the diagram. In diagram 2c, the two
countings, $Q$ and $Q^2$, are for spin-independent and
spin-dependent nucleon polarizabilities contributions,
respectively. }
\end{figure}
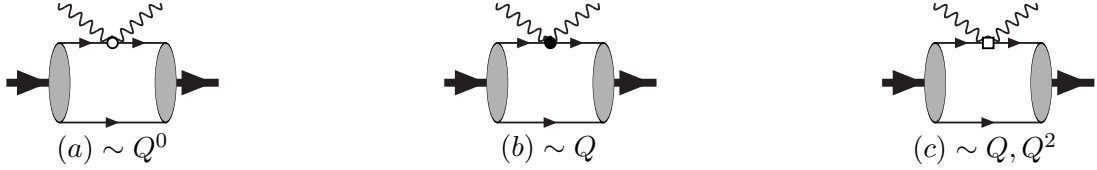

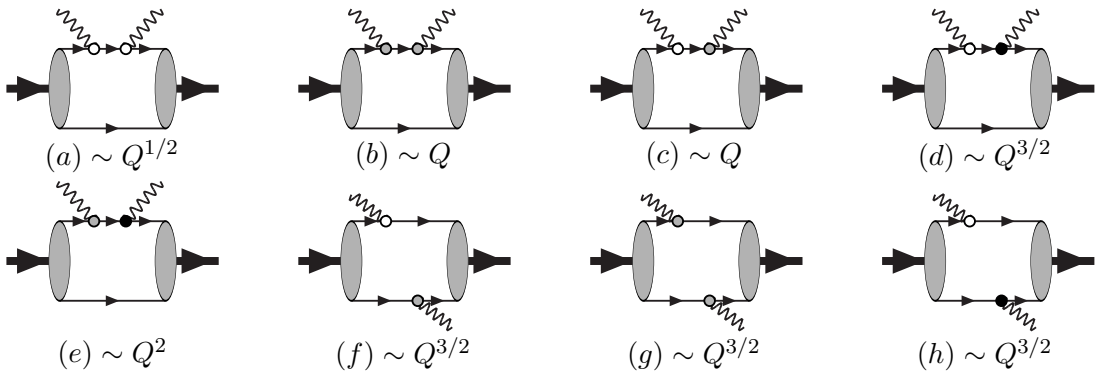
\begin{figure}
\SetWidth{0.7}
\begin{center}
\begin{picture}(420,120)(0,0)
\ArrowLine(20,65)(60,65) \SetWidth{3} \ArrowLine(64,80)(80,80)
\ArrowLine(0,80)(20,80) \SetWidth{0.7} \ArrowLine(20,95)(35,95)
\ArrowLine(35,95)(45,95) \ArrowLine(45,95)(60,95)
\Photon(33,95)(20,110){-2}{5} \Photon(45,95)(58,110){2}{5}
\GCirc(33,95){2}{1} \GCirc(45,95){2}{1}
\GOval(20,80)(15,4)(0){0.7} \GOval(60,80)(15,4)(0){0.7}
\Text(40,55)[]{$(a) \sim Q^{1/2}$} \ArrowLine(130,65)(170,65)
\SetWidth{3} \ArrowLine(174,80)(190,80) \ArrowLine(110,80)(130,80)
\SetWidth{0.7} \ArrowLine(130,95)(145,95)
\ArrowLine(145,95)(155,95) \ArrowLine(155,95)(170,95)
\Photon(143,95)(130,110){-2}{5} \Photon(155,95)(168,110){2}{5}
\GCirc(143,95){2}{.7} \GCirc(155,95){2}{.7}
\GOval(130,80)(15,4)(0){0.7} \GOval(170,80)(15,4)(0){0.7}
\Text(150,55)[]{$(b) \sim Q$} \ArrowLine(240,65)(280,65)
\SetWidth{3} \ArrowLine(284,80)(300,80) \ArrowLine(220,80)(240,80)
\SetWidth{0.7} \ArrowLine(240,95)(255,95)
\ArrowLine(255,95)(265,95) \ArrowLine(265,95)(280,95)
\Photon(253,95)(240,110){-2}{5} \Photon(265,95)(278,110){2}{5}
\GCirc(253,95){2}{1} \GCirc(265,95){2}{.7}
\GOval(240,80)(15,4)(0){0.7} \GOval(280,80)(15,4)(0){0.7}
\Text(260,55)[]{$(c) \sim Q$} \ArrowLine(350,65)(390,65)
\SetWidth{3} \ArrowLine(394,80)(410,80) \ArrowLine(330,80)(350,80)
\SetWidth{0.7} \ArrowLine(350,95)(365,95)
\ArrowLine(365,95)(375,95) \ArrowLine(375,95)(390,95)
\Photon(363,95)(350,110){-2}{5} \Photon(375,95)(388,110){2}{5}
\GCirc(363,95){2}{1} \GCirc(375,95){2}{0}
\GOval(350,80)(15,4)(0){0.7} \GOval(390,80)(15,4)(0){0.7}
\Text(370,55)[]{$(d) \sim Q^{3/2}$} \ArrowLine(20,0)(60,0)
\SetWidth{3} \ArrowLine(64,15)(80,15) \ArrowLine(0,15)(20,15)
\SetWidth{0.7} \ArrowLine(20,30)(35,30) \ArrowLine(35,30)(45,30)
\ArrowLine(45,30)(60,30) \Photon(33,30)(20,45){-2}{5}
\Photon(45,30)(58,45){2}{5} \GCirc(33,30){2}{.7}
\GCirc(45,30){2}{0} \GOval(20,15)(15,4)(0){0.7}
\GOval(60,15)(15,4)(0){0.7} \Text(40,-20)[]{$(e) \sim Q^2$}

\ArrowLine(130,0)(155,0) \SetWidth{3} \ArrowLine(174,15)(190,15)
\ArrowLine(110,15)(130,15) \SetWidth{0.7}
\ArrowLine(130,30)(145,30) \ArrowLine(145,30)(170,30)
\ArrowLine(155,0)(170,0) \Photon(143,30)(130,40){-2}{5}
\Photon(155,0)(168,-10){2}{5} \GCirc(143,30){2}{1}
\GCirc(155,0){2}{.7} \GOval(130,15)(15,4)(0){0.7}
\GOval(170,15)(15,4)(0){0.7} \Text(150,-20)[]{$(f) \sim Q^{3/2}$}
\ArrowLine(240,0)(265,0) \SetWidth{3} \ArrowLine(284,15)(300,15)
\ArrowLine(220,15)(240,15) \SetWidth{0.7}
\ArrowLine(240,30)(255,30) \ArrowLine(255,30)(280,30)
\ArrowLine(265,0)(280,0) \Photon(253,30)(240,40){-2}{5}
\Photon(265,0)(278,-10){2}{5} \GCirc(253,30){2}{.7}
\GCirc(265,0){2}{.7} \GOval(240,15)(15,4)(0){0.7}
\GOval(280,15)(15,4)(0){0.7} \Text(260,-20)[]{$(g) \sim Q^{3/2}$}
\ArrowLine(350,0)(375,0) \SetWidth{3} \ArrowLine(394,15)(410,15)
\ArrowLine(330,15)(350,15) \SetWidth{0.7}
\ArrowLine(350,30)(365,30) \ArrowLine(365,30)(390,30)
\ArrowLine(375,0)(390,0) \Photon(363,30)(350,40){-2}{5}
\Photon(375,0)(388,-10){2}{5} \GCirc(363,30){2}{1}
\GCirc(375,0){2}{0} \GOval(350,15)(15,4)(0){0.7}
\GOval(390,15)(15,4)(0){0.7} \Text(370,-20)[]{$(h) \sim Q^{3/2}$}
\end{picture}
\end{center}
\caption{\it Diagrams without intermediate dibaryons. The small
open circles denote the electric photon-nucleon coupling from the
gauged derivative in the first term in Eq. (\ref{basic}), the
small shaded circles denote the magnetic photon-nucleon coupling
in Eq. (\ref{magnetic}), while the small solid circles represent
the spin-orbit interaction between photon and nucleon in Eq.
(\ref{SO}). }
\end{figure}

%%%%%%%%%%%%%%%%%%%%%%%%%%%%%%%%%
%% FIGURE 4
%%%%%%%%%%%%%%%%%%%%%%%%%%%%%%%%%
\begin{figure}[tbp]
\includegraphics[width=5.6 in]{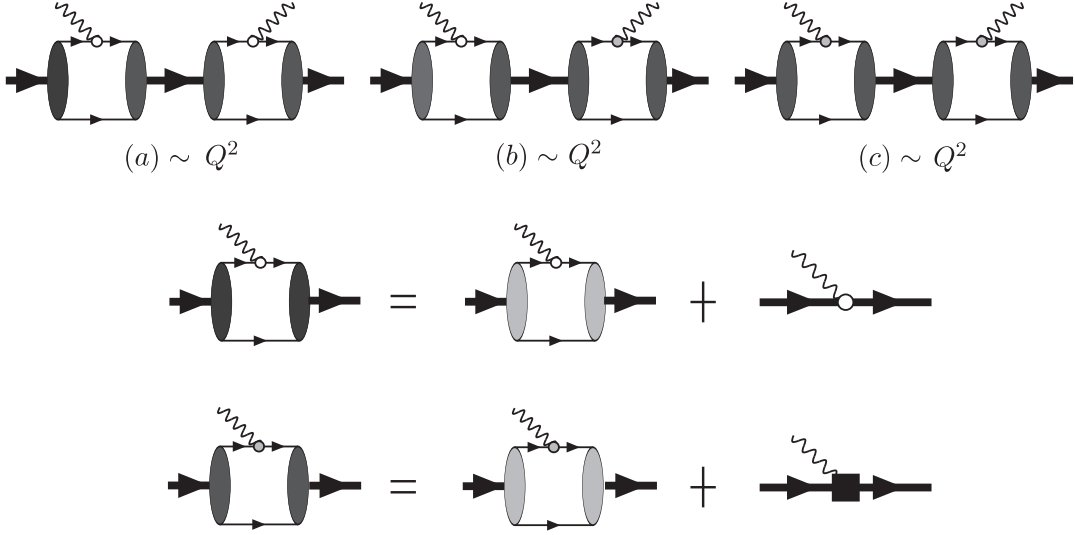}
\caption{\it Diagrams with intermediate dibaryon states. The small
open circles denote the electric coupling in Eq.
(\protect\ref{basic}), and the small shaded circles denote the
magnetic coupling in Eq. (\protect\ref{magnetic}). The
intermediate thick lines with one arrow represent both the spin
singlet and triplet channels. The solid box denotes the $L_1$ and
$L_2$ couplings in Eqs. (\protect\ref{L1}) and (\protect\ref{L2}).
} \label{fig:one}
\end{figure}

To estimate the importance of a particular diagram in our
power-counting scheme, we need to study the dominant regions of a
loop momentum in the integral. Let us use $(q^0, \vec{q})$ to
denote the loop momentum generically. The size of the loop
momentum is determined by poles of the propagators. Typical
nucleon propagators in the loop integration are
${i}/\left({-B-q_0-\frac{\vec{q}^2}{2M}+i\varepsilon}\right)$ when
the photon momentum does not pass through the nucleon line, and
${i}/\left({q_0+\omega-\frac{\vec{q}^2}{2M}+i\varepsilon}\right)$
when the photon momentum does. Because $q_0$ scales as
$|\vec{q}|^2$, the former has a momentum pole at $|\vec{q}| \sim
\sqrt{B}$ and the latter a pole at $|\vec{q}| \sim
\sqrt{\omega}=\sqrt{|\vec{k}|}$. In Region I, these two poles have
the same order of magnitude and have power counting $|\vec{q}|
\sim Q$. In Region II, the pole ($|\vec{q}| \sim \sqrt{B}$) has
counting $|\vec{q}| \sim Q$, and the other pole ($|\vec{q}| \sim
\sqrt{\omega}$) has $|\vec{q}| \sim \sqrt{Q}$. A Feynman integral
can be approximated by the pole that produces a leading
contribution.

For example, let us count the power of diagram (b) in Fig. 3. The
Feynman integral has a momentum power
$Q\omega^2{|\vec{q}|^5}/({|\vec{q}|^6\omega})$, where $Q$ is from
the wave function renormalization, $\omega^2$ is from two magnetic
couplings, the $\omega$ in the denominator is from the propagator
${i}/\left({q_0+\omega-\frac{\vec{q}^2}{2M}+i\varepsilon}\right)$,
and $|\vec{q}|$ is the loop momentum, with $d^4q$ counted as
$|\vec{q}|^5$ and three other propagators in the denominator as
$|\vec{q}|^6$. At the pole $|\vec{q}|\sim Q$, it is of order $Q$;
and at the other pole $|\vec{q}| \sim \sqrt{Q}$, it is $Q^{3/2}$.
Thus the leading contribution is of order $Q$, shown below the
diagram. Note that the $Q$-counting here is dimensionally balanced
by the nucleon mass $M_N$ in the denominator.

Because there are multiple leading regions in a Feynman diagram,
power counting can be rather tricky sometimes. For example, the
nominally higher-order, spin-orbit couplings can produce leading
contributions in a certain momentum region. To see this, let us
compare the power counting for diagrams (f) and (h) in Fig. 3. The
counting of (f) is
$Q{|\vec{q}|^5(\vec{q}+\vec{k})\omega}/(|\vec{q}|^4\omega^2)$,
where in the denominator $|\vec{q}|^4$ is from the two propagators
that do not depend on the photon momentum and $\omega^2$ is from
two propagators that do; in the numerator $(\vec{q}+\vec{k})$ and
$\omega$ factors are from the derivative and magnetic couplings,
respectively. Since only the $\vec{k}$ term in $(\vec{q}+\vec{k})$
survives the symmetrical momentum integration, diagram (f) is of
order $Q^{3/2}$. On the other hand, counting of diagram (h) is
$Q{|\vec{q}|^5(\vec{q}+\vec{k})^2\omega}/({|\vec{q}|^4\omega^2})$
which, compared to diagram (f), has an extra power of
$(\vec{k}+\vec{q})/M_N$, because it is a relativistic correction.
However, the dominant term contributing to the integral is
$\vec{q}^2$ in the $(\vec{q}+\vec{k})^2$ factor, which is of order
$Q$ at the leading pole. Therefore, diagram (h) is also of order
$Q^{3/2}$. Thus the spin-orbit coupling contributes as
significantly as the magnetic coupling in these diagrams.

Power counting allows us to determine the leading contribution of
every Feynman diagram. The result is indicated below each diagram
in Figs. 1, 2, 3, and 4. Again, all countings so far are in terms
of powers of $Q/M_N$, including that for the nucleon
polarizability in diagram 2c. We will treat them as
$(Q/\Lambda)^2$ in phenomenology as we mentioned in the beginning
of this section. According to chiral perturbation theory, the
spin-independent polarizabilties contribute to the scalar
amplitudes at order $\left({Q}/{\Lambda} \right)^2$ \cite{rupak},
the spin-dependent ones contribute to the vector amplitudes at
order $\left({Q}/{\Lambda} \right)^4$. An explanation of the
counting of the nucleon polarizability contributions from diagram
2c is in order. Compared with the leading order contribution T2a
in Appendix B, the result of T2c is suppressed by $2M_N (\alpha_0
\omega^2,\beta_0 \omega^2,\gamma^{(s)}_{E1,M1,E2,M2}\omega^3)
/\alpha_{em}$, which is numerically $(Q/\Lambda)^2$ for scalar
polarizabilties and $(Q/\Lambda)^4$ for vector ones.

According to the above, the scalar amplitudes start at
$\left({Q}/{\Lambda} \right)^0$, vector amplitudes at
$\left({Q}/{\Lambda} \right)^2$ and tensor amplitudes at $\left(
{Q}/{\Lambda} \right)^3$. However, the leading-order vector
amplitudes are actually proportional to the square of the
isovector magnetic moment $\mu_1^2$, and are numerically larger
than what simple power counting indicates. Therefore, the
vector-vector contribution to the unpolarized cross section is
quite significant \cite{unpolarized}. On the other hand, the
enhancement makes the contribution of nucleon spin-dependent
polarizabilties relatively less important.

From Figs. 1 to 4, the vector-polarized amplitudes can be
calculated to order $\left({Q}/{\Lambda} \right)^4$. Our explicit
results are shown in Appendix B. In order to have the result look
more compact, the integration over Feymann parameter $x$ has not
been completed. One must exercise caution, however, when the power
of the un-integrated result is counted. For example, the result of
diagram (b) in Fig. 3 seems to scale as $\gamma\omega^2/(M_N
\omega)^{3/2} \sim Q^{3/2}$. However, after the integration, it
actually scales as $Q$, consistent with power counting.

\section{A Double-Helicity Dependent (Vector-Polarized) Cross Section}

With the scalar and vector amplitudes presented in the previous
section and Appendix B, we can calculate spin-dependent Compton
scattering cross sections. Of course, any polarized cross section
can be constructed out of the complete 12 (scalar, vector, and
tensor) amplitudes once they are known. Because the tensor
amplitudes start at order $(Q/\Lambda)^3$, we do not need to know
them to predict certain spin-dependent cross sections up to some
orders in $Q$.

As we have seen in the previous section, the vector amplitudes
receive contribution from the spin-dependent polarizabilities of
the nucleon. Therefore, we would like to find a cross section
which can be used to probe the vector amplitudes, and hence
possibly extract the spin polarizabilities.

A double-helicity dependent cross section satisfies the above
condition. Suppose the helicities of the initial-state photon and
deuteron are $\lambda_1$ and $\Lambda_1$, respectively. The
general Compton scattering cross section with these polarized
initial states is $\sigma_{\lambda\Lambda}$. Define a
vector-polarized cross section
\begin{equation}
     \Delta_1 \sigma = \frac{1}{2}\left(\sigma_{+1-1}-
     \sigma_{+1+1}\right) ~ ,
     \label{vec cross}
\end{equation}
where $+1$ $(-1)$ is a right-handed (left-handed) polarization. If
the initial momentum of the photon is along the $z$ direction, the
scattered photon momentum is taken along a direction with a polar
angle $\theta$. Then the polarization vector of the in-coming
photon is ${\bf e}=-\frac{i}{\sqrt{2}}({\bf \hat{x}} + i{\bf
\hat{y}})$. The deuteron, moving in the negative $z$ direction
with a negative helicity, has the same wave function. The deuteron
with a positive helicity has a wave function ${\bf \xi} =
\frac{i}{\sqrt{2}}({\bf \hat{x}} - i{\bf \hat{y}})$. Note that the
beam is circularly-polarized in the so defined vector-polarized
cross section. Actually, investigations indicate that the vector
amplitudes cannot be probed as leading order contributions if the
beam is parallel-polarized.

\begin{figure}[t]
\begin{center}
\mbox{
\begin{picture}(0,340)(250,0)
\put(0,190){\insertfig{8.3}{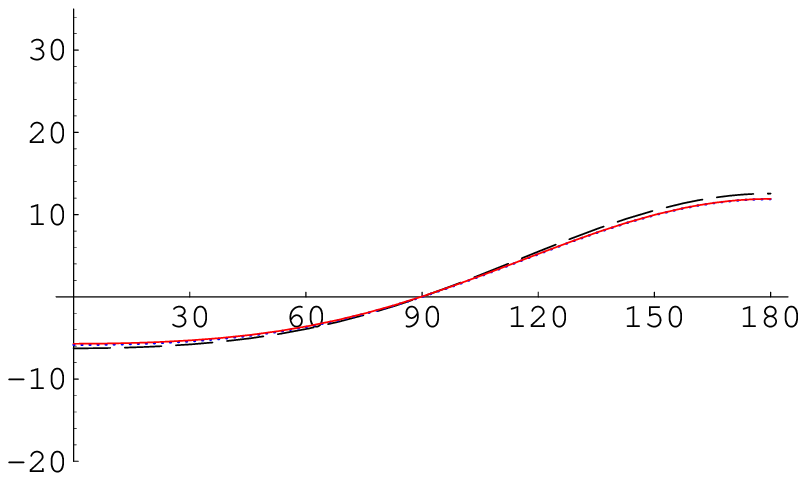}}
\put(260,190){\insertfig{8.3}{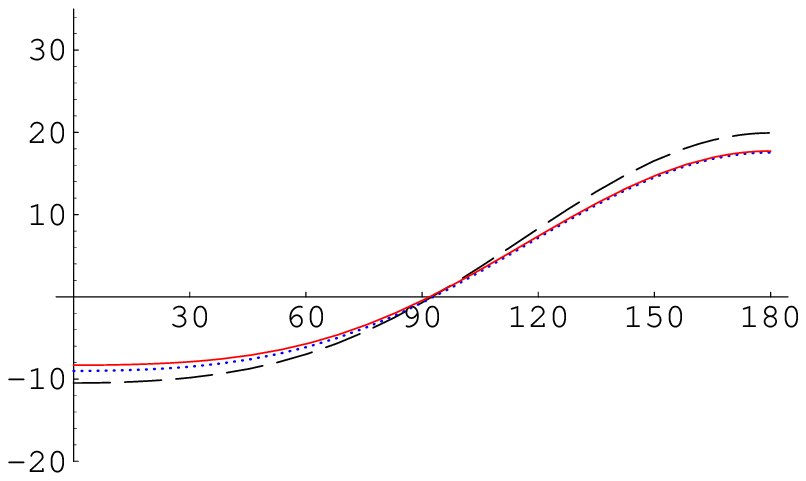}}
\put(0,0){\insertfig{8.3}{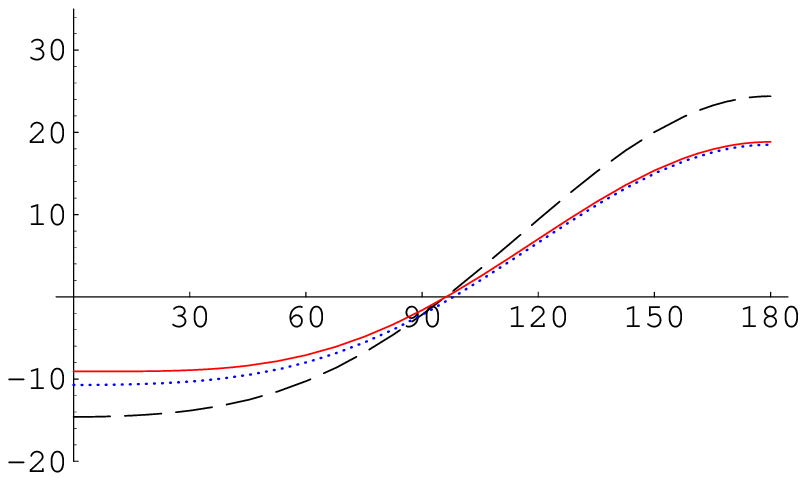}} \put(260,0){\insertfig{8.3}{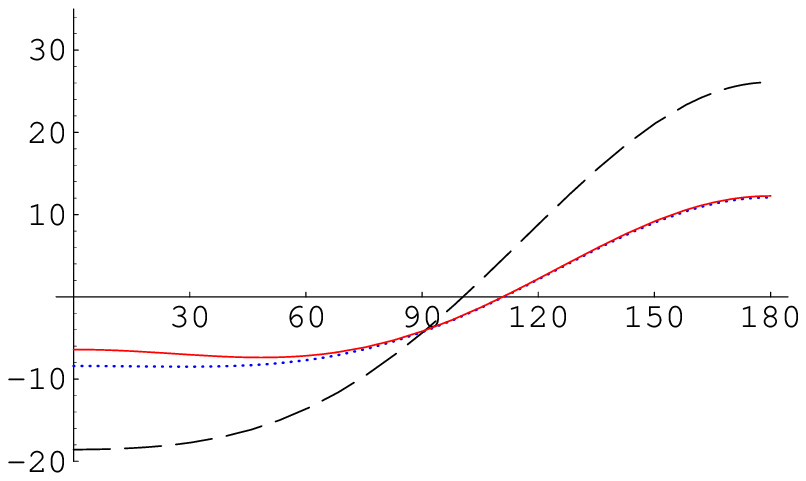}}

\Text(230,40)[c]{$\theta$}\Text(490,40)[c]{$\theta$}
\Text(230,230)[c]{$\theta$}\Text(490,230)[c]{$\theta$}

\Text(45,140)[c]{$\Delta_1
\sigma(\rm{nb})$}\Text(305,140)[c]{$\Delta_1 \sigma(\rm{nb})$}
\Text(45,330)[c]{$\Delta_1
\sigma(\rm{nb})$}\Text(305,330)[c]{$\Delta_1 \sigma(\rm{nb})$}

\Text(120,100)[c]{\large{$\omega=70$
MeV}}\Text(380,100)[c]{\large{$\omega=90$ MeV}}
\Text(120,290)[c]{\large{$\omega=30$
MeV}}\Text(380,290)[c]{\large{$\omega=50$ MeV}}

\end{picture}
}
\end{center}
\caption{\it The vector-polarized cross sections for different CM
frame photon energy $\omega = 30,50,70,90$ {\rm MeV}. (See text
for comments on the 70 and 90 MeV cases). $\theta$ is the
scattering angle in the CM frame. The dashed lines contain no
contribution from spin-independent or spin-dependent
polarizabilities. The dotted lines have contributions from
spin-independent polarizabilities of the nucleon, but without
dependent ones of nucleon. The solid lines have contribution from
both. The values of nucleon polarizabilities are taken from chiral
perturbation theory in Eq. (\ref{pol value}). }
\end{figure}

According to the above definition, the vector-polarized Compton
cross section can be expressed in terms of the full 12 amplitudes
as follows,
\begin{eqnarray}
&&\Delta_1 \sigma ={\rm Re}\left[ S^\ast V + V^\ast V + V^\ast T + T^\ast T \right]\\
                    &=&\frac{2M^2_N}{3(\omega+\sqrt{\omega^2+M^2_D})^2} {\rm Re}\left[[-6(1+z^2)(f_1^*f_3+f_2^*f_4)-12z(f_1^*f_4+f_2^*f_3)
                    \right.\nonumber \\ &&
                    -6z(3+z^2)(f_1^*f_5+f_2^*f_6)
-6(1+3z^2)(f_1^*f_6+f_2^*f_5)]
                \nonumber \\ &&
+[-3(1-z^2)(f_3^*f_3+f_4^*f_4+f_3^*f_6+f_4^*f_5) \nonumber \\
                    &&-3z(1-z^2)(f_3^*f_5+f_4^*f_6)]+[-4(2-z^2)(f_3^*f_7+f_4^*f_8)-4z(f_3^*f_8+f_4^*f_7)\nonumber \\
                    &&-5z(1-z^2)(f_3^*f_9+f_4^*f_{10})-5(1-z^2)(f_3^*f_{10}+f_4^*f_9)
                    +(1+7z^2)(f_3^*f_{11}+f_4^*f_{12}) \nonumber \\
                    &&+z(5+3z^2)(f_3^*f_{12}+f_4^*f_{11})
                    -z(9-z^2)(f_5^*f_7+f_6^*f_8)-(5+3z^2)(f_5^*f_8+f_6^*f_7)  \nonumber \\
                    &&-2(1-z^4)(f_5^*f_9+f_6^*f_{10})
                    -4z(1-z^2)(f_5^*f_{10}+f_6^*f_9)+2z(3+5z^2)(f_5^*f_{11}+f_6^*f_{12})  \nonumber \\
                    &&+(1+12z^2+3z^4)(f_5^*f_{12}+f_6^*f_{11})]+[-3(3+z^2)(f_7^*f_7+f_8^*f_8)+24zf_7^*f_8 \nonumber \\
                    &&+9z(1-z^2)(f_7^*f_9+f_8^*f_{10})-15(1-z^2)(f_7^*f_{10}+f_8^*f_9)
                    +3(1-z^2)(f_7^*f_{11}+f_8^*f_{12}) \nonumber \\
                    &&+3z(1-z^2)(f_7^*f_{12}+f_8^*f_{11})
                    -6(1-z^2)^2(f_9^*f_9+f_{10}^*f_{10}) \nonumber \\
                    && \left. +6z(1-z^2)(f_9^*f_{11}+f_{10}^*f_{12})
                    +3(1-z^4)(f_9^*f_{12}+f_{10}^*f_{11})]\right]
                    ~,
\end{eqnarray}
where $S^\ast V$, $V^\ast V$, $V^\ast T$, and $T^\ast T$ denote
combinations of scalar-vector, vector-vector, vector-tensor, and
tensor-tensor amplitudes, respectively. According to power
counting, the dominant contribution is from the scalar and vector
interference, and is of order $(Q/\Lambda)^2$. If calculating the
cross section to order $(Q/\Lambda)^4$, we need the scalar
amplitudes to order $(Q/\Lambda)^2$ and vector amplitudes to
$(Q/\Lambda)^4$, including the nucleon polarizability term. The
tensor amplitudes do not contribute at this order. Therefore, the
vector-polarized cross section is a useful observable to probe the
vector amplitudes, and hence the spin polarizabilities.

We have shown in Fig. 5 the vector-polarized cross section to $(Q
/ \Lambda)^4$ in EFT at CM photon energy $\omega = 30, 50, 70, 90$
MeV, respectively. The contribution from spin-independent
polarizabilities of the nucleon is more significant at higher
energy. There is virtually no difference between the cross
sections with the polarizabilities turned on or off at the photon
energy $\omega = 30$ MeV. However, there is a notable difference
at $50$ MeV and substantial difference at $70$ MeV and $90$ MeV.
[Note, however, that our results for 70 and 90 MeV are just for
exploratory study, because the pion has to be included as a
dynamical degree of freedom at such high energies. However, we
expect that the general features will not change in a full
analysis.] The effect of the nucleon polarizabilities is more
significant at forward and backward angles (almost zero at
$\theta=\pi/2$). Moreover, the contribution from spin-independent
polarizabilities $\alpha_0,\beta_0$ is of similar size at forward
and backward angles, while the spin-dependent polarizabilities
contribute mainly at forward angles.

According to power counting, both the scalar and spin
polarizabilities contribute to the vector-polarized cross section
at order $\left(Q/ \Lambda \right)^4$. However, the leading-order
vector amplitude is enhanced by a factor $\mu_1^2$. Therefore, the
scalar polarizabilities contribute more significantly to the cross
section, and generate a larger influence than the spin
polarizabilities. As seen in the figure,
$\Delta_1\sigma$---especially at the backward angles---is very
sensitive, as is the unpolarized cross section, to the scalar
polarizabilities of the nucleon. Therefore one cannot extract the
vector polarizabilities without knowing the scalar ones to a
reasonable accuracy. From Fig. 5, the best way to extract the spin
polarizabilities is to measure $\Delta_1 \sigma$ at forward angles
and at relatively high energy (higher than 50 MeV). On the other
hand, the EFT expansion becomes less reliable at high energy.

\section{Asymmetries Sensitive to spin-independent nucleon polarizabilities}

As seen from the previous section, the spin-independent nucleon
polarizabilities have to be determined before the extraction of
spin-dependent ones become possible. In this section, we
investigate various asymmetries with the goal of extracting
spin-independent polarizabilities.

Asymmetries are generally easier to measure than cross sections
because of the cancellation of systematic errors. The asymmetry
associated with the vector-polarized cross section in the previous
section is:
\begin{equation}
\Sigma_z=\frac{ \sigma_{+1-1}-\sigma_{+1+1} }{
\sigma_{+1-1}+\sigma_{+1+1} }.
\end{equation}
where the indices $\pm1$ have the same meaning as in Eq. (\ref{vec
cross}). The expression for the numerator has been shown in the
previous section. The expression for the denominator in terms of
scalar and vector amplitudes is:
\begin{eqnarray}
&& \frac{1}{2}\left( \sigma_{+1-1}+\sigma_{+1+1} \right)
\nonumber \\
 &=& \frac{2M^2_N}{(\omega+\sqrt{\omega^2+M^2_D})^2}
{\rm Re} \left[ (1+z^2)(f^*_1 f_1 + f^*_2 f_2) + 4z(f^*_1
f_2+f^*_3 f_4) + 2(f^*_3 f_3 + f^*_4 f_4)  \right. \nonumber \\
&&  + \frac{1}{2}(3+12z^2+z^4)(f^*_5 f_5 + f^*_6 f_6) +
2z(5+3z^2)f^*_5 f_6 + (3+5z^2)(f^*_3 f_6 + f^*_4 f_5) \nonumber
\\
&& \left. + z(7+z^2)(f^*_3 f_5 + f^*_4 f_6)
 \right]
\end{eqnarray}
The result of $\Sigma_z$ for CM photon energy $\omega$ = 30, 50
MeV is shown in Fig. 6. Clearly, as the vector-polarized cross
section, the asymmetry at the backward angle has stronger
dependence on $\alpha_0$, $\beta_0$ compared to other angles and
shows almost no sensitivity on spin-dependent polarizabilities.
However, unlike the cross section, the dependence on the
$\alpha_0$, $\beta_0$ in the asymmetry is suppressed to about
$8\%$ at 50 MeV due to cancellation between the numerator and
denominator.

\begin{figure}[t]
\begin{center}
\mbox{
\begin{picture}(0,150)(250,0)

\put(0,0){\insertfig{8.3}{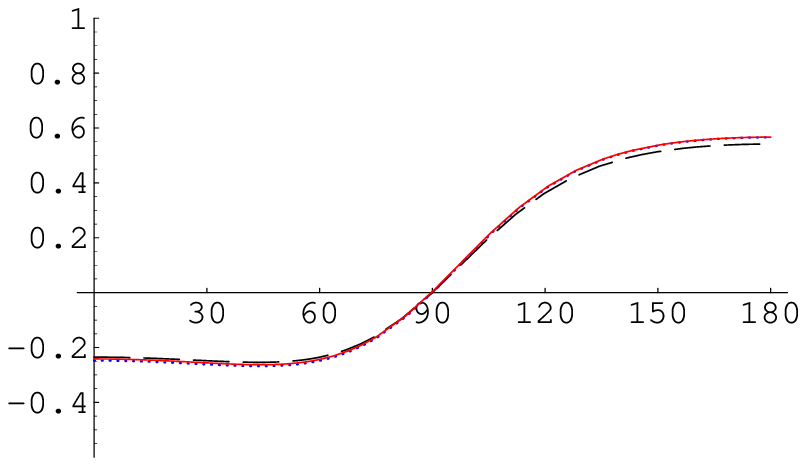}}
\put(260,0){\insertfig{8.3}{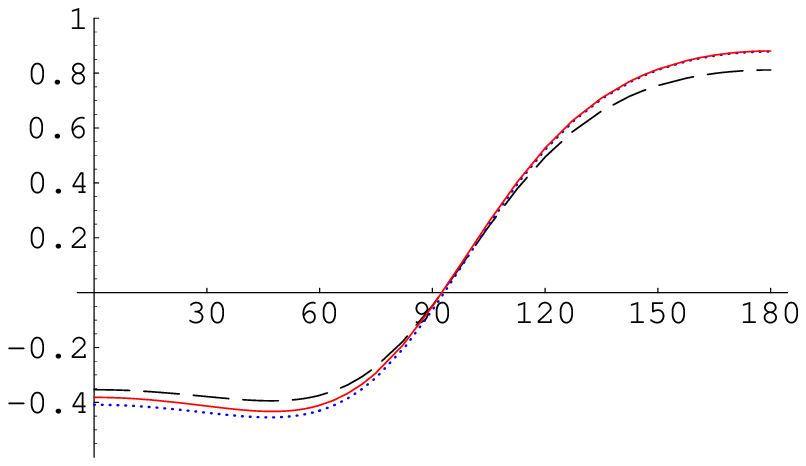}}

\Text(230,40)[c]{$\theta$}\Text(490,40)[c]{$\theta$}

\Text(37,140)[c]{$\Sigma_z$}\Text(297,140)[c]{$\Sigma_z$}

\Text(120,100)[c]{\large{$\omega=30$
MeV}}\Text(380,100)[c]{\large{$\omega=50$ MeV}}

\end{picture}
}
\end{center}
\caption{\it The asymmetry $\Sigma_z$ for different CM frame
photon energy $\omega = 30,50$ {\rm MeV}. $\theta$ is the
scattering angle in the CM frame. The meaning of dashed lines,
dotted lines, and solid lines are the same as in Fig. 5. The
values of nucleon polarizabilities are taken from chiral
perturbation theory in Eq. (\ref{pol value}). }
\end{figure}

\begin{figure}[t]
\begin{center}
\mbox{
\begin{picture}(0,150)(250,0)

\put(0,0){\insertfig{8.3}{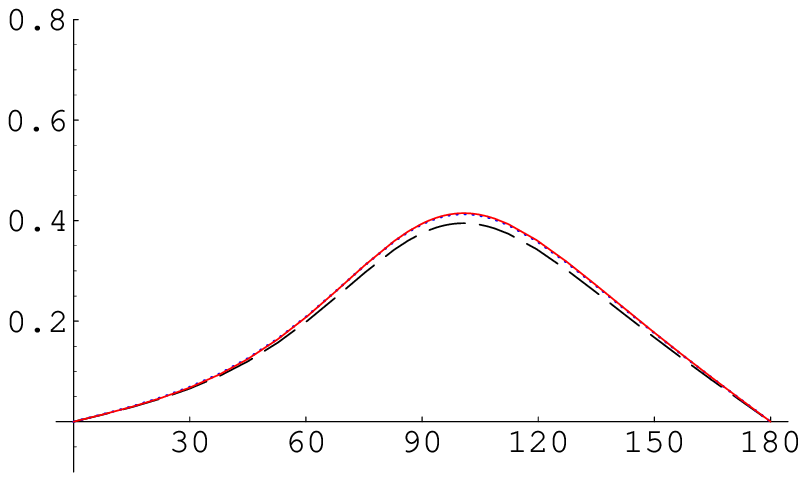}}
\put(260,0){\insertfig{8.3}{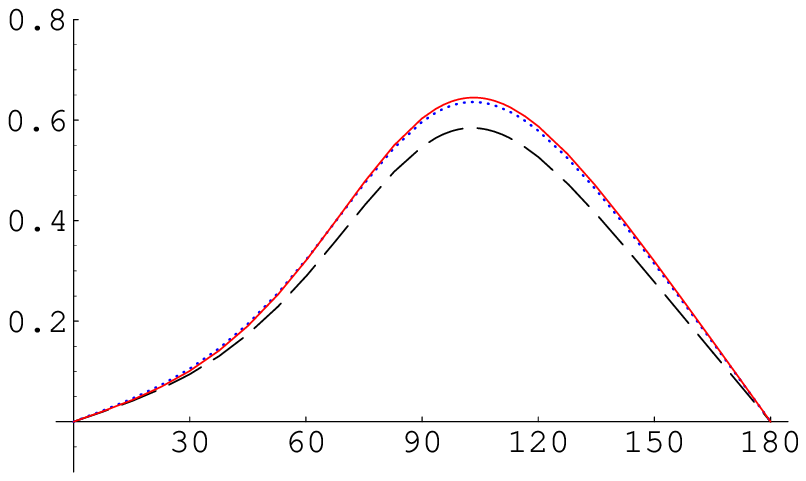}}

\Text(230,30)[c]{$\theta$}\Text(490,30)[c]{$\theta$}

\Text(37,140)[c]{$-\Sigma_x$}\Text(297,140)[c]{$-\Sigma_x$}

\Text(120,90)[c]{\large{$\omega=30$
MeV}}\Text(380,125)[c]{\large{$\omega=50$ MeV}}

\end{picture}
}
\end{center}
\caption{\it The asymmetry $\Sigma_x$ for different CM frame
photon energy $\omega = 30,50$ {\rm MeV}. $\theta$ is the
scattering angle in the CM frame. The meaning of dashed lines,
dotted lines, and solid lines are the same as in Fig. 5. The
values of nucleon polarizabilities are taken from chiral
perturbation theory in Eq. (\ref{pol value}). }
\end{figure}

\begin{figure}[t]
\begin{center}
\mbox{
\begin{picture}(0,150)(250,0)

\put(0,0){\insertfig{8.3}{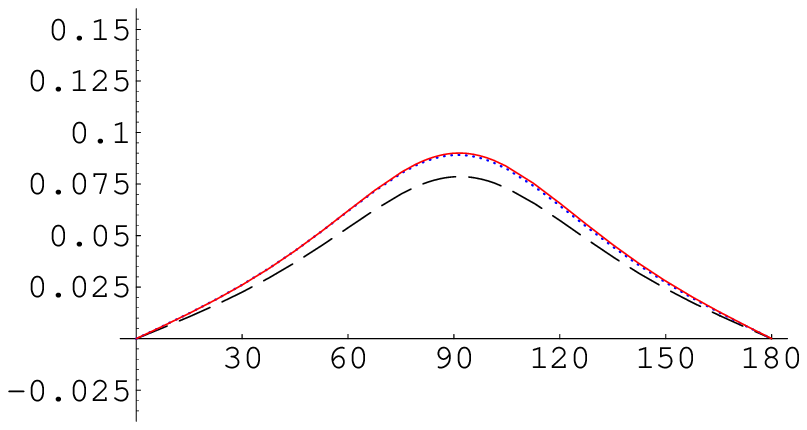}}
\put(260,0){\insertfig{8.3}{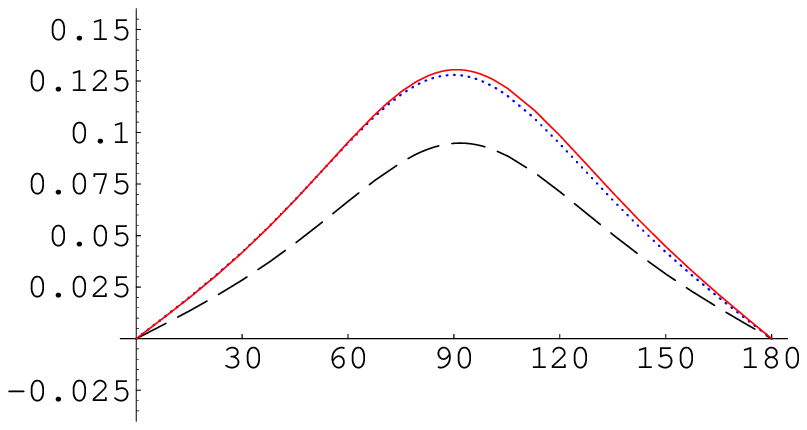}}

\Text(230,45)[c]{$\theta$}\Text(490,45)[c]{$\theta$}

\Text(50,135)[c]{$\Sigma_y$}\Text(310,135)[c]{$\Sigma_y$}

\Text(120,100)[c]{\large{$\omega=30$
MeV}}\Text(380,125)[c]{\large{$\omega=50$ MeV}}

\end{picture}
}
\end{center}
\caption{\it The asymmetry $\Sigma_y$ for different CM frame
photon energy $\omega = 30,50$ {\rm MeV}. $\theta$ is the
scattering angle in the CM frame. The meaning of dashed lines,
dotted lines, and solid lines are the same as in Fig. 5. The
values of nucleon polarizabilities are taken from chiral
perturbation theory in Eq. (\ref{pol value}). }
\end{figure}

In the following, we investigate other asymmetries in search of a
larger dependence on $\alpha_0$, $\beta_0$. There are two new
asymmetries related to $\Sigma_z$ when the polarization axis of
the deuteron target is changed. If the $xz$ plane is chosen as the
scattering plane, one can define an asymmetry with deuteron
polarized linearly in the $x$ direction:
\begin{equation}
\Sigma_x=\frac{\sigma_{+1,J_x=+1}-\sigma_{+1,J_x=-1}}{\sigma_{+1,J_x=+1}+\sigma_{+1,J_x=-1}},
\end{equation}
with the first index +1 of $\sigma$ indicating that the photon is
right-handed polarized, the second index indicating that the
deuteron target is polarized in the $J_x=\pm 1$ states. The
expressions for the numerator and the denominator in terms of
scalar and vector amplitudes are:
\begin{eqnarray}
&&\frac{1}{2}\left( \sigma_{+1,J_x=+1}-\sigma_{+1,J_x=-1} \right)
\nonumber \\
&=& \frac{2M^2_N}{(\omega+\sqrt{\omega^2+M^2_D})^2} \sqrt{1-z^2}
{\rm Re} \left[ -2\left( z f^*_1 f_3 + f^*_1 f_4 + (1+z^2)f^*_1
f_5 +
2z f^*_1 f_6 + f^*_2 f_3  \right. \right. \nonumber \\
&& \left.  + z f^*_2 f_4 + 2zf^*_2 f_5 +
(1+z^2) f^*_2 f_6 \right) + z(f^*_3 f_3 + f^*_4 f_4) + 2f^*_3 f_4 + (1+z^2) f^*_3 f_5 + 2zf^*_3 f_6  \nonumber \\
&& \left. + 2zf^*_4 f_5 + (1+z^2) f^*_4 f_6 \right] \nonumber \\
&& \frac{1}{2}\left( \sigma_{+1,J_x=+1}+\sigma_{+1,J_x=-1} \right)
\nonumber \\
 &=& \frac{2M^2_N}{(\omega+\sqrt{\omega^2+M^2_D})^2}
{\rm Re} \left[ (1+z^2)(f^*_1 f_1 + f^*_2 f_2) + 4zf^*_1 f_2 +
(2-z^2)(f^*_3 f_3 + f^*_4 f_4)  \right. \nonumber \\
&& \left. + 2zf^*_3 f_4 + (3+z^2)(f^*_3 f_6 + f^*_4 f_5) +
z(5-z^2)(f^*_3 f_5 + f^*_4 f_6)  \right. \nonumber \\
&& \left. + \frac{1}{2}(3+6z^2-z^4)(f^*_5 f_5 + f^*_6 f_6) + 8z
f^*_5 f_6 \right]
\end{eqnarray}
The result for $\Sigma_x$ at CM photon energy $\omega$ = 30, 50
MeV is shown in Fig. 7. The peak of this asymmetry is around the
scattering angle of 105 degrees, where the dependence on
$\alpha_0$, $\beta_0$ is about $8\%$ at 50 MeV.

Similarly, one can define the asymmetry with the deuteron
polarized in the $y$ direction, which is perpendicular to the
scattering plane. It turns out this asymmetry is actually a
single-spin asymmetry, independent on the polarization of the
photon beam.
\begin{equation}
\Sigma_y=\frac{\sigma_{J_y=+1}-\sigma_{J_y=-1}}{\sigma_{J_y=+1}+\sigma_{J_y=-1}}
\ ,
\end{equation}
where the photon beam is unpolarized and the deuteron target is
polarized in the $J_y= \pm 1$ states. The expressions for the
numerator and the denominator in terms of scalar and vector
amplitudes are:
\begin{eqnarray}
\frac{1}{2}\left( \sigma_{J_y=+1}-\sigma_{J_y=-1} \right) &=& -i
\frac{4M^2_N \sqrt{1-z^2}}{(\omega+\sqrt{\omega^2+M^2_D})^2}
 {\rm Im} \left[ f^*_1 f_4 + f^*_2 f_3 + z f^*_1 f_3 +
z f^*_2 f_4 \right] \nonumber \\
\frac{1}{2}\left( \sigma_{J_y=+1}+\sigma_{J_y=-1} \right) &=&
\frac{4M^2_N}{(\omega+\sqrt{\omega^2+M^2_D})^2} {\rm Re} \left[
\frac{1}{2}( 1+z^2 ) \left( f^*_1f_1 + f^*_2f_2
\right) \right. \nonumber \\
&&  + \frac{1}{2}(2-z^2)\left( f^*_3f_3 + f^*_4f_4 \right)
+\frac{1}{2}(1+3z^2)\left( f^*_5f_5 + f^*_6f_6 \right) \nonumber
\\
&& + (1+z^2)\left( f^*_3f_6 + f^*_4f_5 \right) + 2z\left( f^*_1f_2
+ f^*_3f_5 + f^*_4f_6 \right) \nonumber \\
&& + zf^*_3f_4 + z(3+z^2) f^*_5f_6
\end{eqnarray}
The result for $\Sigma_y$ at CM photon energy $\omega$ = 30, 50
MeV is shown in Fig. 8. The peak of this asymmetry is around a
scattering angle of 90 degree, where the dependence on $\alpha_0$,
$\beta_0$ is about $12\%$ at 30 MeV and $26\%$ at 50 MeV,  much
larger than the dependence in $\Sigma_{y,z}$. Therefore, the
single-spin asymmetry should serve as a good observable to extract
nucleon scalar-isoscalar polarizabilities. Note that the
polarizations of the deuteron in the above asymmetries are defined
in the CM frame, while in experiment the deuteron is prepared
polarized in the lab frame. The polarization in these two frames
are different in case of $\Sigma_{x,y}$. This is an error of size
$\omega/M_D$ which can be safely neglected at low energy.

The tensor amplitudes contributions are not taken into account in
these results shown above. They are small contributions from the
analysis of the power counting. But numerically, the effect of
them could be enhanced due to the large size of isovector nucleon
magnetic moment, which also explains that the vector amplitudes
effect are enhanced. While a more complete calculation of
asymmetries with all the tensor amplitudes included is beyond the
scope of this paper, we did, however, study their effects on the
asymmetries by using the tensor amplitude $f_7$ from a previous
calculation in EFT with pion \cite{tensor}. We found that the
$\Sigma_y$ is less dependent on these amplitudes compared with the
other asymmetries, which offers an additional reason that this
asymmetry is better than others for extracting $\alpha_0$ and
$\beta_0$.

We have also investigated the parallel-perpendicular single spin
asymmetry, which is the ratio of the difference and sum of two
cross sections when the deuteron target is unpolarized and the
photon beam is linearly polarized either parallel or perpendicular
to the scattering plane. This asymmetry is found to have a weak
dependence (about $3\%$ at 50 MeV) on $\alpha_0$, $\beta_0$ than
$\Sigma_{x,y,z}$ and therefore is not presented here.

\section{conclusion}

In this paper, we presented a convenient set of basis for Compton
scattering on the deuteron. We then calculated the scalar and
vector Compton amplitudes to ${\cal O}((Q/\Lambda)^4)$ in a
nuclear EFT without the pion, at which the scalar and spin
polarizabilities of the nucleon contribute. The result was then
used to calculate a double-helicity-dependent cross section which
is linearly proportional to the vector amplitudes. We studied the
effects of the polarizabilities on the cross section, finding that
the scalar polarizabilities have more dominant influence than the
spin polarizabilities. Thus an accurate measurement of the cross
section can help to determine the former. However, if the scalar
polarizabilities are determined with good accuracy, the cross
section can provide a constraint on the spin-dependent ones.
Finally, we investigated various asymmetries in search of large
dependence on scalar polarizabilities and found that $\Sigma_y$
has the best potential.

This work was supported by the U. S. Department of Energy via
grant DE-FG02-93ER-40762 and by the National Science Council of
Taiwan, ROC. JWC thanks Paulo Bedaque for organizing the Summer
Lattice Workshop 2004 at Lawrence Berkeley Laboratory where part
of this research was completed.

\appendix
\section{Tensor Basis for Deuteron Compton Amplitudes}

The 12 basis structures can be systematically obtained by keeping
track of the matrix structure sandwiched between the initial and
final deuteron polarization states. The structures with unit
matrix and single spin matrix are the same as the structures for
spin-1/2 target. There are six such structures ($\rho_1 \sim
\rho_6$) \cite{babusci}. Our goal is to find out the remaining six
structures, which should all be of tensor type with symmetrized
double spin matrixes.

To write them down, first notice that since there are double $J$s
associated with them, the parity invariance requires that there
are an even number of cross products among vectors:
$\hat{\epsilon},\hat{\epsilon}'^{*},\hat{k},\hat{k}'$ and two
$J$s. Moreover, since any even number of cross products can be
transformed into dot products, we only need to write down
structures with dot products. Since subtracting trace is
straightforward, we choose to do it at the end. The structures
before subtracting trace can be found systematically by looking at
which pair dot with $J$s and what is left over. First, if the pair
is $\hat{\epsilon}$ and $\hat{\epsilon}'^{*}$, There is only one
such structure:
\begin{equation}
\tau_1=J \cdot \hat{\epsilon} J \cdot \hat{\epsilon}'^{*} + J
\cdot \hat{\epsilon}'^{*} J \cdot \hat{\epsilon} ~ .
\end{equation}
If the pair is $\hat{k}$ and $\hat{k}'$, there are two structures:
\begin{eqnarray}
\tau_2 &=& \hat{\epsilon}'^{*} \cdot \hat{\epsilon} (J \cdot
\hat{k} J \cdot \hat{k}' + J \cdot \hat{k}' J \cdot \hat{k})~,
\nonumber \\
\tau_3 &=& \hat{k} \cdot \hat{\epsilon}'^{*} \hat{k}' \cdot
\hat{\epsilon}(J \cdot \hat{k} J \cdot \hat{k}' + J \cdot \hat{k}'
J \cdot \hat{k}) ~ .
\end{eqnarray}
If the pair is $\hat{k}$ and $\hat{\epsilon}'^{*}$, time reversal
invariance requires that the other pair $\hat{k}'$ and
$\hat{\epsilon}$ appear in the same structure and in the following
combination:
\begin{equation}
\tau_{4}=\hat{\epsilon}'^{*} \cdot \hat{k} (J \cdot \hat{\epsilon}
J \cdot \hat{k}' + J \cdot \hat{k}' J \cdot
\hat{\epsilon})+\hat{\epsilon} \cdot \hat{k}' (J \cdot
\hat{\epsilon}'^{*} J \cdot \hat{k} + J \cdot \hat{k} J \cdot
\hat{\epsilon}'^{*}) ~ .
\end{equation}
If the pair is $\hat{k}$ and $\hat{\epsilon}$, time reversal
invariance requires that the other pair, $\hat{k}'$ and
$\hat{\epsilon}'^{*}$, appear in the same structure and in the
following combination:
\begin{equation}
\tau_{5}=\hat{\epsilon}'^{*} \cdot \hat{k} (J \cdot \hat{\epsilon}
J \cdot \hat{k} + J \cdot \hat{k} J \cdot
\hat{\epsilon})+\hat{\epsilon} \cdot \hat{k}' (J \cdot
\hat{\epsilon}'^{*} J \cdot \hat{k}' + J \cdot \hat{k}' J \cdot
\hat{\epsilon}'^{*}) ~ .
\end{equation}
If the pair is two $\hat{k}$s, the time reversal invariance
requires that the other pair, two $\hat{k}'$s, appears in the same
structure and in the proper combination. There are two structures
of this type:
\begin{eqnarray}
\tau_{6} &=& \hat{\epsilon}'^{*} \cdot \hat{\epsilon} (J \cdot
\hat{k} J \cdot \hat{k} + J \cdot \hat{k}' J \cdot \hat{k}')~ ,
\nonumber \\
\tau_{7} &=& \hat{k} \cdot \hat{\epsilon}'^{*} \hat{k}' \cdot
\hat{\epsilon}(J \cdot \hat{k} J \cdot \hat{k} + J \cdot \hat{k}'
J \cdot \hat{k}') ~ .
\end{eqnarray}

The above way of constructing structures with double $J$s exhausts
all possibilities. There is no problem about the completeness.
However, we get more structures than expected from helicity
counting. It is hard to find the relation among them directly and
it turns out that we need to make use of the duality character of
the electric magnetic field. Starting from the above seven
structures, we can write down another set of structures which
covers the above set and has the duality correspondence among
them, just like the structures from $\rho_1$ to $\rho_6$. Without
knowing the dependence among the structures from $\tau_1$ to
$\tau_{7}$, the minimal number of such a set of structures is
eight. They are chosen as:
\begin{eqnarray}
\tau_1' &=& J \cdot \hat{\epsilon} J \cdot \hat{\epsilon}'^{*} + J
\cdot
\hat{\epsilon}'^{*} J \cdot \hat{\epsilon}~, \nonumber \\
\tau_2'&=& J \cdot \hat{s} J \cdot \hat{s}'^{*} + J \cdot
\hat{s}'^{*} J
\cdot \hat{s} ~, \nonumber \\
\tau_3' &=& \hat{\epsilon}'^{*} \cdot \hat{\epsilon} (J \cdot
\hat{k} J \cdot \hat{k}' + J \cdot \hat{k}' J \cdot \hat{k})~,
\nonumber \\
\tau_4' &=& \hat{s}'^{*} \cdot \hat{s} (J \cdot \hat{k} J \cdot
\hat{k}' + J \cdot \hat{k}' J \cdot \hat{k})~,
\nonumber \\
\tau_5' &=& \hat{\epsilon}'^{*} \cdot \hat{k} (J \cdot
\hat{\epsilon} J \cdot \hat{k}' + J \cdot \hat{k}' J \cdot
\hat{\epsilon})+\hat{\epsilon} \cdot \hat{k}' (J \cdot
\hat{\epsilon}'^{*} J \cdot \hat{k} + J \cdot
\hat{k} J \cdot \hat{\epsilon}'^{*})~, \nonumber \\
\tau_6' &=& \hat{s}'^{*} \cdot \hat{k} (J \cdot \hat{s} J \cdot
\hat{k}' + J \cdot \hat{k}' J \cdot \hat{s})+\hat{s} \cdot
\hat{k}' (J \cdot \hat{s}'^{*} J \cdot \hat{k} + J \cdot \hat{k} J
\cdot \hat{s}'^{*})~,
\nonumber \\
\tau_7' &=& \hat{\epsilon}'^{*} \cdot \hat{\epsilon} (J \cdot
\hat{k} J \cdot \hat{k} + J \cdot \hat{k}' J \cdot \hat{k}')~,
\nonumber \\
\tau_8' &=& \hat{s}'^{*} \cdot \hat{s} (J \cdot \hat{k} J \cdot
\hat{k} + J \cdot \hat{k}' J \cdot \hat{k}')~.
\end{eqnarray}
One notices that under duality transformation, these eight
structures transform as: $\tau_{2i-1}' \Leftrightarrow
\tau_{2i}'$, with $i=1,2,3,4$. This set with eight structures can
be expressed in term of seven $f_i$s and the expression is found
to be:
\begin{eqnarray}
&&\tau_1'=\tau_1~, ~ ~ \tau_2'=4\rho_2 + \tau_4- z\tau_1 -
\tau_2~, ~ ~ \tau_3'= \tau_2~, ~ ~
\tau_4'=z\tau_2-\tau_3~, \nonumber \\
&&\tau_5'=\tau_4~, ~ ~ \tau_6'=2z\tau_2+\tau_5-2\tau_3-2\tau_6~, ~
~ \tau_7'=\tau_6~, ~ ~ \tau_8=z\tau_6 - \tau_7~.
\end{eqnarray}
Since eight structures are expressed in terms of other seven
structures, one relation among $\tau_i'$s must exist, and it is
found to be:
\begin{equation}
z\tau_1'+\tau_2'+\tau_3'-\tau_5'=4\rho_2~,
\end{equation}
from which another relation can be found through duality
transformation of the above relation:
\begin{equation}
z\tau_2'+\tau_1'+\tau_4'-\tau_6'=4\rho_1~.
\end{equation}
Now, we have two constraints on eight structures and are therefore
left with six independent structures, as expected from helicity
counting. We choose $\tau_{1,2,5,6,7,8}'$ as the basis structures.
With trace subtracted explicitly, they are:
\begin{eqnarray}
\rho_7&=&-(J \cdot \hat{\epsilon} J \cdot \hat{\epsilon}'^{*} + J
\cdot \hat{\epsilon}'^{*} J \cdot
\hat{\epsilon}-\frac{4}{3}\hat{\epsilon}'^{*} \cdot
\hat{\epsilon})=\hat{\xi}'^{*} \cdot \hat{\epsilon}'^{*} \hat{\xi}
\cdot \hat{\epsilon} + \hat{\xi}'^{*} \cdot \hat{\epsilon}
\hat{\xi} \cdot \hat{\epsilon}'^{*} -\frac{2}{3}\hat{\xi}'^{*}
\cdot
\hat{\xi}\hat{\epsilon}'^{*} \cdot \hat{\epsilon} ~, \nonumber \\
\rho_8&=&-(J \cdot \hat{s} J \cdot \hat{s}'^{*} + J \cdot
\hat{s}'^{*} J \cdot \hat{s}-\frac{4}{3}\hat{s}'^{*} \cdot
\hat{s})=\hat{\xi}'^{*} \cdot \hat{s}'^{*} \hat{\xi} \cdot \hat{s}
+ \hat{\xi}'^{*} \cdot \hat{s} \hat{\xi} \cdot \hat{s}'^{*}
-\frac{2}{3}\hat{\xi}'^{*} \cdot
\hat{\xi}\hat{s}'^{*} \cdot \hat{s}~, \nonumber \\
\rho_9&=&-\hat{\epsilon}'^{*} \cdot \hat{k} (J \cdot
\hat{\epsilon} J \cdot \hat{k}' + J \cdot \hat{k}' J \cdot
\hat{\epsilon})-\hat{\epsilon} \cdot \hat{k}' (J \cdot
\hat{\epsilon}'^{*} J \cdot \hat{k} + J \cdot
\hat{k} J \cdot \hat{\epsilon}'^{*})+\frac{8}{3}\hat{\epsilon}'^{*} \cdot \hat{k} \hat{\epsilon} \cdot \hat{k}' \nonumber \\
&=&\hat{\epsilon}'^{*} \cdot \hat{k} (\hat{\xi}'^{*} \cdot
\hat{k}' \hat{\xi} \cdot \hat{\epsilon}
 + \hat{\xi}'^{*} \cdot \hat{\epsilon} \hat{\xi} \cdot \hat{k}') + \hat{\epsilon} \cdot \hat{k}'
 (\hat{\xi}'^{*} \cdot \hat{k} \hat{\xi} \cdot \hat{\epsilon}'^{*}
 + \hat{\xi}'^{*} \cdot \hat{\epsilon}'^{*} \hat{\xi} \cdot \hat{k}) -
 \frac{4}{3}\hat{\xi}'^{*} \cdot \hat{\xi} \hat{\epsilon}'^{*} \cdot \hat{k} \hat{\epsilon} \cdot \hat{k}'~, \nonumber \\
\rho_{10}&=&-\hat{s}'^{*} \cdot \hat{k} (J \cdot \hat{s} J \cdot
\hat{k}' + J \cdot \hat{k}' J \cdot \hat{s})-\hat{s} \cdot
\hat{k}' (J \cdot \hat{s}'^{*} J \cdot \hat{k} + J \cdot
\hat{k} J \cdot \hat{s}'^{*})+\frac{8}{3}\hat{s}'^{*} \cdot \hat{k} \hat{s} \cdot \hat{k}' \nonumber \\
&=&\hat{s}'^{*} \cdot \hat{k} (\hat{\xi}'^{*} \cdot \hat{k}'
\hat{\xi} \cdot \hat{s}
 + \hat{\xi}'^{*} \cdot \hat{s} \hat{\xi} \cdot \hat{k}') + \hat{s} \cdot \hat{k}'
 (\hat{\xi}'^{*} \cdot \hat{k} \hat{\xi} \cdot \hat{s}'^{*}
 + \hat{\xi}'^{*} \cdot \hat{s}'^{*} \hat{\xi} \cdot \hat{k}) -
 \frac{4}{3}\hat{\xi}'^{*} \cdot \hat{\xi} \hat{s}'^{*} \cdot \hat{k} \hat{s} \cdot \hat{k}'~, \nonumber \\
\rho_{11}&=&-\hat{\epsilon}'^{*} \cdot \hat{\epsilon} (J \cdot
\hat{k} J \cdot \hat{k} + J \cdot \hat{k}' J \cdot \hat{k}'
-\frac{4}{3})=\hat{\epsilon}'^{*} \cdot
\hat{\epsilon}(\hat{\xi}'^{*} \cdot \hat{k}\hat{\xi} \cdot \hat{k}
+ \hat{\xi}'^{*} \cdot \hat{k}'\hat{\xi} \cdot \hat{k}'
-\frac{2}{3} \hat{\xi}'^{*} \cdot \hat{\xi})~,
\nonumber \\
\rho_{12}&=&-\hat{s}'^{*} \cdot \hat{s} (J \cdot \hat{k} J \cdot
\hat{k} + J \cdot \hat{k}' J \cdot \hat{k}'
-\frac{4}{3})=\hat{s}'^{*} \cdot \hat{s}(\hat{\xi}'^{*} \cdot
\hat{k}\hat{\xi} \cdot \hat{k} + \hat{\xi}'^{*} \cdot
\hat{k}'\hat{\xi} \cdot \hat{k}' - \frac{2}{3}\hat{\xi}'^{*} \cdot
\hat{\xi})~.
\end{eqnarray}

$\rho_i$s ($i=1 \sim 12$) are the basis structures of deuteron
Compton amplitudes in the frame where time reversal invariance is
manifest such as the Breit frame and center-of-mass frame. Note
that lab frame is not such a frame because it lacks the symmetry
between the initial and final deuteron.

There are other tensor structures that are often met in studies of
Compton scattering on the deuteron. Here we provide a list and
their relation to the basis set defined above:
\begin{eqnarray}
&&\hat{\epsilon}'^{*} \cdot \hat{\epsilon}( \hat{\xi}'^{*} \cdot
\hat{k}\hat{\xi} \cdot \hat{k}' + \hat{\xi}'^{*} \cdot
\hat{k}'\hat{\xi} \cdot \hat{k} ) = -z\rho_7
-\rho_8 + \rho_9 + \frac{2}{3}z \rho_1 ~, \nonumber \\
&&\hat{\epsilon}'^{*} \cdot \hat{k} \hat{\epsilon} \cdot \hat{k}'(
\hat{\xi}'^{*} \cdot \hat{k}\hat{\xi} \cdot \hat{k}' +
\hat{\xi}'^{*} \cdot \hat{k}'\hat{\xi} \cdot \hat{k} ) =
(1-z^2)\rho_7 + z\rho_9 - \rho_{10} + \frac{2}{3}z^2\rho_1 -
\frac{2}{3}z\rho_2 ~, \nonumber \\
&&\hat{\epsilon}'^{*} \cdot \hat{k} \hat{\epsilon} \cdot \hat{k}'(
\hat{\xi}'^{*} \cdot \hat{k}\hat{\xi} \cdot \hat{k} +
\hat{\xi}'^{*} \cdot \hat{k}'\hat{\xi} \cdot \hat{k}' ) =
z\rho_{11} - \rho_{12} + \frac{2}{3}(z\rho_1 -
\rho_2) ~, \nonumber \\
&&\hat{\epsilon}'^{*} \cdot \hat{k}( \hat{\xi}'^{*} \cdot
\hat{\epsilon}\hat{\xi} \cdot \hat{k} + \hat{\xi}'^{*} \cdot
\hat{k}\hat{\xi} \cdot \hat{\epsilon} ) + \hat{\epsilon} \cdot
\hat{k}'( \hat{\xi}'^{*} \cdot \hat{\epsilon}'^{*} \hat{\xi} \cdot
\hat{k}' + \hat{\xi}'^{*} \cdot \hat{k}' \hat{\xi} \cdot
\hat{\epsilon}'^{*} ) = 2\rho_7 + 2z\rho_8 -\rho_{10}+2\rho_{11}~,
\nonumber \\
&&\hat{\epsilon}'^{*} \cdot \hat{\epsilon}(\hat{\xi}'^{*} \times
\hat{\xi}) \cdot ({\hat{k}' \times \hat{k}}) = z\rho_3 + \rho_4 -
\rho_5 ~, \nonumber \\
&&(\hat{\xi}'^{*} \times \hat{\xi}) \cdot (\hat{\epsilon}'^{*}
\times \hat{k}) \hat{\epsilon} \cdot \hat{k}' - (\hat{\xi}'^{*}
\times \hat{\xi}) \cdot (\hat{\epsilon} \times \hat{k}')
\hat{\epsilon}'^{*} \cdot \hat{k} = 2z\rho_3 - \rho_5  ~,\nonumber
\\
&&(\hat{\xi}'^{*} \times \hat{\xi}) \cdot \hat{s}
\hat{\epsilon}'^{*} \cdot \hat{k} - (\hat{\xi}'^{*} \times
\hat{\xi}) \cdot \hat{s}'^{*} \hat{\epsilon} \cdot \hat{k}' =
2\rho_3 - \rho_6 ~ ,
\end{eqnarray}
where $z=\cos\theta=\hat{k} \cdot \hat{k}'$, which is used
throughout this paper. The last three expressions for the
vector-type structures have appeared in the literature before
\cite{babusci}. Other useful relations can be obtained from the
above through the duality transformation.

\section{Compton amplitudes to $(Q/\Lambda)^4$ in EFT}

Diagrams with the photon directly coupled to the dibaryon are
shown in Fig. 1. The result is:

\begin{eqnarray}
T_{1a} &=& i \frac{e^2}{2M_N} \frac{\gamma r^{(3S_1)}}{1-\gamma r^{(3S_1)}} \rho_1 ~, \nonumber \\
T_{1c} &=& i \frac{e^2}{4M^2_N} \frac{\gamma r^{(3S_1)}}{1-\gamma
r^{(3S_1)}} \omega \rho_3 (1-4\mu_0 + \frac{4L_2}{r^{(3S_1)}}) ~.
\end{eqnarray}

Diagrams with the seagull interaction on the nucleon line are
shown in Fig. 2, among which are contributions from nucleon
polarizabilities. The result for each diagram is:
\begin{eqnarray}
T_{2a} &=& -i \frac{4e^2}{M_N} \frac{\gamma}{1-\gamma r^{(3S_1)}}
\frac{1}{\omega \sqrt{2-2z}}
\rm{arctan}(\frac{\omega\sqrt{2-2z}}{4\gamma}) \rho_1 ~, \nonumber \\
T_{2b} &=& i \frac{2e^2}{M_N^2} \frac{\gamma}{1-\gamma r^{(3S_1)}}
\left( (2\mu_0-\frac{1}{2})+(2\mu_1-\frac{1}{2})) \right)
 \frac{1}{\sqrt{2-2z}}
\rm{arctan}(\frac{\omega\sqrt{2-2z}}{4\gamma}) \rho_3 ~, \nonumber \\
T_{2c} &=& i 32\pi  \frac{\gamma}{1-\gamma r^{(3S_1)}} \frac{1}{
\sqrt{2-2z}} \rm{arctan}(\frac{\omega\sqrt{2-2z}}{4\gamma}) \left[
\alpha_0 \omega \rho_1 + \beta_0 \omega \rho_2 - \gamma_{E1}
\omega^2 \rho_3 - \gamma_{M1} \omega^2 \rho_4  \right. \nonumber \\
&& \left. + \gamma_{E2} \omega^2 (-\rho_4 + \rho_5) + \gamma_{M2}
\omega^2 (\rho_3 - \rho_6) \right] ~,
\end{eqnarray}
with $T_{2c}$ associated with the nucleon polarizabilities.

The contribution without the intermediate singlet or triplet state
is from diagrams in Fig. 3. The result of each diagram along with
photon crossing and the diagram with interchange of two photon
coupling vertices, if different, is:
\begin{eqnarray}
T_{3a} &=&i\frac{e^{2}}{2M_{N}}\frac{\gamma }{1-\gamma
r^{(3S_{1})}}\left[ \rho _{1}\left(
\int_{0}^{1}dx\frac{1-x}{\sqrt{\gamma ^{2}+M_{N}\omega x-i\epsilon
}}+\int_{0}^{1}dx\frac{1-x}{\sqrt{\gamma ^{2}-M_{N}\omega
x-i\epsilon }}\right) \right.   \nonumber \\
&&\left. +\omega ^{2}(z\rho _{1}-\rho _{2})\left( \frac{1}{4}\int_{0}^{1}dx%
\frac{\frac{1}{6}(1-x)^{3}+(1-x)(2-x)}{(\gamma ^{2}+M_{N}\omega
x-i\epsilon
)^{3/2}}\right. \right.   \nonumber \\
&&\left. \left.
+\frac{1}{24}\int_{0}^{1}dx\frac{(1-x)^{3}}{(\gamma
^{2}-M_{N}\omega x-i\epsilon )^{3/2}}\right) \right] ~,  \nonumber \\
T_{3b} &=& -i \frac{e^2}{4M_N} \frac{\gamma}{1-\gamma r^{(3S_1)}}
\left( \mu^2_0+\mu^2_1 \right) \omega^2 \nonumber \\
&& \times \left[ (\rho_4-\rho_2) \int^1_0 dx \frac{1-x}{(\gamma^2
+ M_N\omega x -i\epsilon)^{3/2}}
 - (\rho_4+\rho_2)\int^1_0 dx \frac{1-x}{(\gamma^2
- M_N\omega x -i\epsilon)^{3/2}} \right] ~, \nonumber \\
T_{3c} &=& i \frac{e^2}{16M_N} \frac{\gamma}{1-\gamma r^{(3S_1)}}
\left( \mu_0 + \mu_1 \right) \omega^2 \left( \rho_6-2\rho_3
\right) \nonumber \\
&& \times  \left( \int^1_0 dx \frac{x^2-4x+3}{(\gamma^2 +
M_N\omega x -i\epsilon)^{3/2}} - \int^1_0 dx
\frac{(1-x)^2}{(\gamma^2 -
M_N\omega x -i\epsilon)^{3/2}}\right) ~, \nonumber \\
T_{3d} &=& -i \frac{e^2}{4M_N^2} \frac{\gamma}{1-\gamma
r^{(3S_1)}}
\left( 2\mu_0 + 2\mu_1 -1 \right) \omega \rho_3 \nonumber \\
&& \times  \left( \int^1_0 dx \frac{1-x}{\sqrt{\gamma^2 +
M_N\omega x -i\epsilon}}  + \int^1_0 dx \frac{1-x}{\sqrt{\gamma^2
- M_N\omega x -i\epsilon}} \right) ~,
\nonumber \\
T_{3e} &=&i\frac{e^{2}}{16M_{N}^{2}}\frac{\gamma }{1-\gamma r^{(3S_{1})}}%
\left( \mu _{0}(2\mu _{0}-\frac{1}{2})+\mu _{1}(2\mu _{1}-\frac{1}{2}%
)\right) \omega ^{3}  \nonumber \\
&&\times \left[ -\frac{1}{2}(2\rho _{1}+\rho _{6})\int_{0}^{1}dx\frac{%
(1-x)^{2}}{(\gamma ^{2}-M_{N}\omega x-i\epsilon )^{3/2}}+(2\rho
_{1}-\rho
_{6})\int_{0}^{1}dx\frac{\frac{3}{2}-2x+\frac{1}{2}x^{2}}{(\gamma
^{2}+M_{N}\omega x-i\epsilon )^{3/2}}\right.   \nonumber \\
&&\left. +(\rho _{2}+\rho
_{4})\int_{0}^{1}dx\frac{1-x^{2}}{(\gamma
^{2}-M_{N}\omega x-i\epsilon )^{3/2}}+(\rho _{2}-\rho _{4})\int_{0}^{1}dx%
\frac{(1-x)^{2}}{(\gamma ^{2}+M_{N}\omega x-i\epsilon
)^{3/2}}\right] ~,
\nonumber \\
T_{3f} &=& -i \frac{e^2}{4M_N^2} \frac{\gamma}{1-\gamma
r^{(3S_1)}}
\left( \mu_0 - \mu_1 \right) \omega (-2\rho_3 + \rho_6) \nonumber \\
&& \times \left( \int^1_0 dx \frac{x}{\sqrt{\gamma^2 + M_N\omega x
-i\epsilon}} - \int^1_0 dx \frac{1-x}{\sqrt{\gamma^2 - M_N\omega x
-i\epsilon}} \right) ~, \nonumber \\
T_{3g} &=& -i \frac{e^2}{2M_N^2} \frac{\gamma}{1-\gamma
r^{(3S_1)}} \left( \mu^2_0 - \mu^2_1 \right) \omega \left(
\frac{1}{3}\rho_2 -
\rho_8 \right) \nonumber \\
&& \times \left( \int^1_0 dx \frac{1}{\sqrt{\gamma^2 + M_N\omega x
-i\epsilon}} - \int^1_0 dx \frac{1}{\sqrt{\gamma^2 - M_N\omega x
-i\epsilon}} \right) ~, \nonumber \\
T_{3h} &=& i \frac{e^2}{M_N^3} \frac{\gamma}{1-\gamma r^{(3S_1)}}
\left( \mu_0 - \mu_1 \right) \rho_3 \nonumber \\
&& \left( \int^1_0 dx \sqrt{\gamma^2 + M_N\omega x -i\epsilon} -
\int^1_0 dx \sqrt{\gamma^2 - M_N\omega x -i\epsilon} \right) ~.
\end{eqnarray}

The diagrams with the intermediate triplet or singlet state are
shown in Fig. 4. The result from each diagram along with photon
crossing and the diagram with interchange of two photon coupling
vertices, if different, is:
\begin{eqnarray}
T_{4a} &=& i \frac{e^2}{8M_N} \frac{\gamma}{1-\gamma r^{(3S_1)}}
\omega^2
\left(z\rho_1 - \rho_2 \right) \left( \int^1_0 dx \frac{1}{\sqrt{%
\gamma^2+M_N\omega x-i\epsilon}} - r^{(3S_1)} \right)^2  \nonumber \\
&& \times \frac{1}{-\frac{1}{a^{(^3S_1)}}-\frac{1}{2}r^{(3S_1)}(\gamma^2+M_N%
\omega)+\sqrt{\gamma^2+M_N\omega-i\epsilon}} ~,  \nonumber \\
T_{4b} &=&i\frac{e^{2}}{4M_{N}}\frac{\gamma }{1-\gamma
r^{(3S_{1})}}\mu
_{0}\omega ^{2}\left( -2\rho _{3}+\rho _{6}\right) \left( \int_{0}^{1}dx%
\frac{1}{\sqrt{\gamma ^{2}+M_{N}\omega x-i\epsilon
}}-r^{(3S_{1})}\right)
\nonumber \\
&&\times \left( \int_{0}^{1}dx\frac{1}{\sqrt{\gamma
^{2}+M_{N}\omega
x-i\epsilon }}-r^{(3S_{1})}+\frac{L_{2}}{\mu _{0}}\right)   \nonumber \\
&&\times
\frac{1}{-\frac{1}{a^{(^{3}S_{1})}}-\frac{1}{2}r^{(3S_{1})}(\gamma
^{2}+M_{N}\omega )+\sqrt{\gamma ^{2}+M_{N}\omega -i\epsilon }}~, \\
T_{4c} &=& -i \frac{e^2}{4M_N} \frac{\gamma}{1-\gamma r^{(3S_1)}}
\mu^2_0 \omega^2  \nonumber \\
&& \times \left[ \left( -\frac{4}{3}\rho_2 + \rho_4 + \rho_8
\right) \left( \int^1_0 dx \frac{1}{\sqrt{\gamma^2+M_N\omega
x-i\epsilon}} -r^{(3S_1)} +
\frac{L_2}{\mu_0} \right)^2 \right.  \nonumber \\
&& \times \frac{1}{-\frac{1}{a^{(^3S_1)}}-\frac{1}{2}r^{(3S_1)}(\gamma^2+M_N%
\omega)+\sqrt{\gamma^2+M_N\omega-i\epsilon}}  \nonumber \\
&& + \left( -\frac{4}{3}\rho_2 - \rho_4 + \rho_8 \right) \left(
\int^1_0 dx
\frac{1}{\sqrt{\gamma^2-M_N\omega x-i\epsilon}} -r^{(3S_1)} + \frac{L_2}{%
\mu_0} \right)^2  \nonumber \\
&& \times \left. \frac{1}{-\frac{1}{a^{(^3S_1)}}-\frac{1}{2}%
r^{(3S_1)}(\gamma^2-M_N\omega)+\sqrt{\gamma^2-M_N\omega-i\epsilon}}
\right]
  \nonumber \\
&& + i \frac{e^2}{4M_N} \frac{\gamma}{1-\gamma r^{(3S_1)}} \mu^2_1
\omega^2 \left[ \left( \frac{2}{3}\rho_2+\rho_4+\rho_8 \right)
\left( \int^1_0 dx \frac{1}{\sqrt{\gamma^2-M_N\omega x-i\epsilon}} + \frac{%
L_1}{\mu_1} \right)^2 \right.  \nonumber \\
&& \times \frac{1}{-\frac{1}{a^{(^1S_0)}}-\frac{1}{2}r^{(^1S_0)}(%
\gamma^2-M_N\omega)+\sqrt{\gamma^2-M_N\omega-i\epsilon}}  \nonumber \\
&& + \left( \frac{2}{3}\rho_2-\rho_4+\rho_8 \right) \left( \int^1_0 dx \frac{%
1}{\sqrt{\gamma^2+M_N\omega x-i\epsilon}} + \frac{L_1}{\mu_1}
\right)^2
\nonumber \\
&& \times \left. \frac{1}{-\frac{1}{a^{(^1S_0)}}-\frac{1}{2}%
r^{(^1S_0)}(\gamma^2+M_N\omega)+\sqrt{\gamma^2+M_N\omega-i\epsilon}}
\right]
~.  \nonumber \\
&&
\end{eqnarray}

\end{document}